\documentclass[preprint,12pt]{elsarticle}  

\usepackage{amssymb}
\usepackage{amsmath}
\usepackage{amsthm}
\usepackage{bm,amsfonts}
\usepackage{tikz}
\usepackage{pict2e}
\usepackage{verbatim}

\usepackage[a4paper,margin=1in]{geometry}
\usepackage{array}
\usepackage{fontawesome}
\usepackage{tabularx}
\usepackage{multirow}
\usepackage{ragged2e}      % for \RaggedRight in table cells
\usepackage{wasysym}      % for \Lightbulb
\usepackage{enumitem}     % control itemize inside table cells
\renewcommand{\arraystretch}{1.35}
\usepackage{graphicx}
\usepackage{pifont} % for circles

\usepackage{booktabs, hhline}

\usepackage{caption}

\usepackage{xcolor}

\newcommand{\fullcircle}{\ding{108}}   % full black circle
\newcommand{\emptycircle}{\ding{109}}  % empty circle

\newcommand{\rot}[1]{\rotatebox{60}{\parbox{2.2cm}{\centering #1}}}

\newtheorem{example}{Example}

\journal{Computer Networks}

\begin{document}

\begin{frontmatter}



\title{A systematic review of secure coded caching}


\author[RHUL]{S.-L. Ng} 
\author[BBK]{M. B. Paterson}
\author[RHUL]{E. A. Quaglia}
\affiliation[RHUL]{organization={Information Security Group, Royal Holloway, University of London},
            city={Egham},
            postcode={Surrey TW20 0EX}, 
            country={United Kingdom}}
\affiliation[BBK]{organization={School of Computing and Mathematical Science, Birkbeck, University of London},
             addressline={Malet Street, Bloomsbury},
             postcode={London WC1E 7HX},
             country={United Kingdom}}
            
%% Abstract

%% Revision in blue text  {\color{blue}...} - removed for arXiv version

\begin{abstract}
In a content delivery network (CDN), resources are strained during peak-time and underutilised in off-peak times when supplying digital content to users. Caching can help balance this.  At the off-peak time some content is delivered to users' local caches.  During peak time, the use of cached data to serve users' requests relieves strain on the network by reducing repeated transfer of popular content.  In \emph{coded caching}, the cache
content placement is designed in conjunction with the delivery techniques to optimise network
throughput.

Since  dissemination
of  information, as well as the delivery of entertainment, is reliant on CDNs, the security and privacy of cache placement, user demand, and content delivery, are paramount. In much of the literature in \emph{secure coded caching}, security is built on top of solutions that have efficiency in mind, and most current proposals focus on the security of individual parts of the process.  A lack of a unifying network model also makes it difficult to compare or combine solutions.  

In this survey we analyse the security and privacy requirements of secure coded caching, 
and evaluate existing schemes in terms of the security provided and the cost of this security provision.  We also review the techniques used to achieve secure coded caching and analyse their limitations.   In addition, we contextualise secure coded caching in the landscape of other secure content delivery primitives.  As a result, we identify and prioritise open security and privacy challenges for the future.
\footnote{\copyright 2026. This manuscript version is made available under the CC-BY-NC-ND 4.0 license \texttt{https://creativecommons.org/licenses/by-nc-nd/4.0/}}
\end{abstract}


%% Keywords

\end{frontmatter} 


\section{Introduction}
\label{sec:intro}

In supplying digital content to users, network resources are 
strained during peak-time and underutilised in off-peak times.
This can be balanced through {\em caching:} an off-peak {\em
  placement phase} delivers content to users' local caches,
 and a peak-time {\em delivery phase} supplies users' requests
 from the caches or
over the network.  Much early research focussed on the
placement phase.  In {\em coded caching}, an innovation due to
Maddah-Ali and Niesen \cite{MaddahAli2014}, the content placement is
designed in conjunction with the delivery techniques to
optimise network throughput.  This has led to much subsequent
research aimed at developing constructions that provide good trade-offs between various scheme parameters, e.g., \cite{Yan2017, Chittoor2021, Cheng2021}, and/or exploring the limits of such trade-offs under different variants of the originally-studied model, e.g., \cite{MaddahAli2015,Shariatpanahi2016,Karamchandani2016,YanParampalli2017,Lampiris2018,KrishnanNamboodiri2022,KrishnanNamboodiriPeter2022,Das2022}.

Coded caching is motivated by considerations of efficiency.  However,
when such approaches are applied in a real-world context, security
considerations become important. For example, the bug affecting the
Fastly CDN in 2021 caused widespread 
disruption to major websites \cite{fastly}.
Since deployment of government services and dissemination of critical health information, as well as the delivery of entertainment, are reliant on CDNs,
security and privacy are paramount and cannot be an afterthought.  Indeed, CDNs are projected to handle an increasing portion of global internet traffic in the coming years. In 2017, CDNs were responsible for delivering approximately 52\% of internet traffic. This share rose to over 70\% by 2023 \cite{oblivCDN}.
Over the last ten years, various notions of security for coded caching schemes have been studied, starting with the work of Sengupta, Tandon and Clancy \cite{Sengupta2015}.  These include properties such as confidentiality of files against eavesdroppers or unauthorised users, as well as privacy of users' file requests. However, many current proposals build security on top of the solutions that were designed with efficiency in mind, and these proposals largely focus on the security of individual parts of the process.
Different proposals also do not necessarily agree on the same security assumptions in the network model.  

These deficiencies highlight the need for a more systematic approach to allow for comparison of existing schemes on a like-for-like basis and the identification of aspects of security that have not been addressed by existing literature. 
 In this work, we systematise results in the secure coded caching literature, considering the currently accepted variants of network models.
 This allows us to examine and clarify the underlying assumptions regarding the security properties of the server and cache storage and the communication channels, as well as  assumptions regarding the capabilities of the servers, users and potential adversaries. On the basis of this we provide a comprehensive analysis of the range of security challenges inherent in a coded caching environment.  We systematically ana\-lyse the security and privacy requirements of the system and classify existing schemes in terms of the provided security.  In addition, we contextualise secure coded caching in the landscape of other secure primitives designed to deliver data over networks and analyse the extent to which related solutions may be applicable in this environment.  This allows us to identify security and privacy challenges not yet completely addressed in the literature, and to suggest critical directions for future research towards constructing effective whole-system security solutions that meet application needs.

\subsection{Related work}

{
Content caching plays a role in many different types of networks, including CDNs, edge computing and cellular networks, to enable more efficient content delivery.  Security, however, is not always regarded as a priority. Here we review relevant surveys on security in caching. An early survey \cite{Abdullahi2015} discussed caching in information centric networks (ICNs), where content is published and subscribed to by name; hence,  the provision of resources is not seen in the context of an end-to-end connection with a server, but rather as a delivery of named content. The importance of security was recognised in this survey but it did not contain detailed reviews of security mechanisms.  Another early survey \cite{Li2018} discussed caching techniques in cellular networks, and security was discussed generally as a future challenge.  }

{In \cite{Yao2019}, Yao, Han and Ansari give a comprehensive picture of content caching techniques specifically designed for mobile edge computing environments, with discussions on techniques (including coded caching), cache content selection, and cache location.  They discuss aspects of security and review solutions, including encryption against eavesdropping, frequency hopping against jamming, and anonymisation for privacy.  In the context of a CDN, Ghaznavi et al.\ \cite{Ghaznavi2021} discuss caching security  from a networking perspective, such as routing and addressing, and attacks include DoS and cache poisoning.  Pruthvi et al. \cite{Pruthvi2023}  focus on the role of caching in delivering efficient service in ICNs, and security is noted as an issue but not discussed in any depth.  These two surveys (\cite{Ghaznavi2021,Pruthvi2023}) do not consider coded caching explicitly.}

{In more recent surveys, security became more of a focus, and techniques reviewed veered towards a machine learning approach.  The works \cite{Khan2024,Li2024,Nguyen2023} survey content caching techniques (including coded caching) at the mobile network edge, with security discussed as a challenge.  The security techniques they review are predominantly federated learning and noise-based methods to prevent information leakages.   Surveys with security as the main focus include \cite{Zhang2025, XZhang2025}, where caching security is considered from an edge computing network perspective.  The authors of \cite{XZhang2025} focus on the network edge as attack surface, while those of \cite{Zhang2025} focus on the security between layers of the network.  Cache security is maintained through access control and privacy through obfuscation techniques.  }

{In much of the literature, security vulnerabilities are regarded as a disruption of efficiency, and solutions appear to be an afterthought.  In our study, we focus specifically on the security of coded caching - the maintenance of confidentiality of content and privacy of users in conjunction with coding for efficiency.  Table \ref{tab:surveys} summarises this review of related surveys.}

{We note that caching is not limited to passive content.  For example, \cite{Barrios2023} surveys the caching of service codes and service computation at the edge.  The security issues for this application overlaps with that of content caching (for example, the prevention of information leakage), but the suitability of the coding approach of content coded caching for computer codes, and the concerns of code attestation, are beyond the scope of our review.  In another direction, coding methods are use in coded distributed computing (for example, \cite{Ng2021}) to enable a group of computers (``workers'') working together to solve a computational task by sharing networking and storage resources.  Security in this scenario focuses on data privacy against collusion of workers.  However, this model differs from that of secure coded caching in the aims of the distribution and on issues specific to distributed computing (such as stragglers), which again is beyond the scope of our review.}

{We note also that \cite{Naderializadeh2017} implies that caching and multicasting can be done without knowledge of the underlying communication network,  and that a scheme optimal in the bottleneck network \cite{MaddahAli2014} which underlies much of the research in secure coded caching will also achieve optimality within a constant factor for all memoryless networks.  We therefore focus on this model in the remainder of this survey.}
\\

\begin{table}[h!]
\renewcommand{\arraystretch}{1.3}
 {\centering
  \caption{{Summary of related surveys}} 
  \label{tab:surveys}
\begin{tabular}{ p{1.5cm}p{1.5cm}p{2.1cm}p{6.0cm}  }
%\begin{tabular}{ rlp{4.5cm}l }
 \hline
 {Survey (year)} & {Network type} &  {Security discussion} & {Analysis/review of solution}\\
 \hline
 \cite{Abdullahi2015} (2015) & ICN &  Mentioned &  No detailed review\\
 \cite{Li2018} (2018) &  Cellular & Challenge &   No detailed review \\
\cite{Yao2019} (2019) & Edge  & Challenge & Encryption, frequency hopping, anonymisation \\
 \cite{Ghaznavi2021} (2021) & CDN & Main focus &  Network security solutions \\
\cite{Pruthvi2023} (2023) & ICN &  Mentioned &  No detailed review\\
\cite{Khan2024,Li2024,Nguyen2023} (2023-4)  & Edge  & Challenge &  Noise-based, machine learning techniques \\
\cite{Zhang2025,XZhang2025} (2025) & Edge & Main focus & Noise-based, machine learning techniques\\
Our work&  General & Main focus & Analysis of security requirement integral to coded caching\\
 \hline
\end{tabular}  }
\end{table}

\subsection{Paper structure and contributions} 
This paper is structured as follows. Section~\ref{sec:SCCmodel} establishes the model that forms the basis for our systematization.  We begin by describing the underlying architecture and the phases involved in the coded caching process.  In Section~\ref{subsec:assumptionsvariations} we make explicit the various assumptions about the model that are made in the literature, and we point out variations on the model that have been considered.  We highlight general issues in the modelling of security of these schemes in Section~\ref{subsec:securitymodelling}, and these observations inform our subsequent discussion. 

Section~\ref{sec:survey} provides a survey of existing literature on the design of secure coded caching schemes.   We bear in mind that the underlying goal of coded caching is the reduction in peak-time use of the broadcast channel, and in Section~\ref{subsec:approaches} we review how this is impacted by the security techniques that have been proposed in the literature.  We group this literature according to the security properties that are considered.  

 In Section~\ref{subsec:analysisanddiscussion} we give an overall analysis of this literature, identifying commonalities and highlighting key omissions from a security perspective.

Section~\ref{sec:related} gives an overview of various content delivery primitives relating closely to coded caching that have been studied from the point of view of security.  In each case we identify similarities and differences in the underlying model, and discuss the extent to which approaches from this area provide relevant insight to the design of secure coded caching schemes.

As a result of our analysis we identify in Section \ref{Open} a range of directions in which further research will advance the state of the art in secure coded caching.  Section \ref{sec:strongermodel} advocates two directions in which further research may contribute to a stronger and more realistic security model.  Section \ref{sec:moderntechniques} identifies some areas in the provision of security that can benefit from new techniques, and Section \ref{sec:ecosystem} asks questions about how we may clarify the relationship between secure coded caching and other secure content delivery primitives.  
The desired outcome is that these further studies will lead to effective whole-system solutions meeting real-world application needs in terms of efficiency and security.

\section{A model for (secure) coded caching}
\label{sec:SCCmodel}
%\section{A model for coded caching}
%\begin{itemize}
%\item the three phases
%\item assumptions
%\item papers that follow this, and papers that don't and why this is the most commonly accepted
%\item A nice table
%\end{itemize

\begin{figure}[htb]
\centering
\begin{tikzpicture}[scale=0.60]
%\draw[step=1cm,gray,very thin] (0,0) grid (11,10);

%the users
\draw (0,2.5) -- (0,3) -- (2,3) -- (2,2.5) -- (0,2.5);
\node at (1, 2.7) {{\scriptsize $U_0$}};
\draw (3,2.5) -- (3,3) -- (5,3) -- (5,2.5) -- (3,2.5);
\node at (4, 2.7) {{\scriptsize $U_1$}};
\draw (8,2.5) -- (8,3) -- (10,3) -- (10,2.5) -- (8,2.5);
\node at (9, 2.7) {{\scriptsize $U_{K-1}$}};
\node at (11, 2.8) {{\scriptsize $K$ users}};

% user-cache
\draw[thick] (1,2.5) -- (1,2);
\draw[thick] (4,2.5) -- (4,2);
\draw[thick] (9,2.5) -- (9,2);

%the caches
\draw (0,1) -- (0,2) -- (2,2) -- (2,1) -- (0,1);
\node at (1,1.5) {{\scriptsize $C_0$}};
\draw (3,1) -- (3,2) -- (5,2) -- (5,1) -- (3,1);
\node at (4,1.5) {{\scriptsize $C_1$}};
\node at (6.5,2) {$\ldots$};
\draw (8,1) -- (8,2) -- (10,2) -- (10,1) -- (8,1);
\node at (9,1.5) {{\scriptsize $C_{K-1}$}};
\draw[arrows=<->](8,0.7)--(10,0.7);
\node at (9,0.4) {{\scriptsize $n$}};
\draw[arrows=<->](7.8,1)--(7.8,2);
\node at (7.5,1.5) {{\scriptsize $M$}};
\node at (11, 1.5) {{\scriptsize Caches}};

%the server
\draw (4,6) -- (4,9) -- (6,9) -- (6,6) --(4,6);
\draw (4,8.5) -- (6,8.5);
\node at (5,8.7) {{\scriptsize $W_0$}};
\draw (4,8) -- (6,8);
\node at (5,8.2) {{\scriptsize $W_1$}};
\draw (4,6.5) -- (6,6.5);
\node at (5,7.5) {$\vdots$};
\node at (5,6.2) {{\scriptsize $W_{N-1}$}};
\draw[arrows=<->](4,9.2)--(6,9.2);
\node at (5,9.5) {{\scriptsize $n$}};
\draw[arrows=<->](6.2,6)--(6.2,9);
\node at (6.5,7.5) {{\scriptsize $N$}};
\node at (3.3, 8) {{\scriptsize Server}};
\node at (3.3, 7.5) {{\scriptsize $S$}};

%broadcast channel
\draw[thick,arrows=->] (5,6) -- (5,5.5);
\draw[thick] (5,5.5) -- (5,5);
\draw[thick,arrows=->] (5,5) -- (1,3);
\draw[thick,arrows=->] (5,5) -- (4,3);
\draw[thick,arrows=->] (5,5) -- (9,3);
\node at (6.6,5.4) {{\scriptsize $B_{(d_0, \ldots, d_{K-1})}$}};

%demands
%\draw[thick,dashed,arrows=->] (1,3.2) -- (2,4);
%\node at (1.2,3.7) {$d_0$};
%\draw[thick,dashed,arrows=->] (3.7,3.2) -- (4.2,4);
%\node at (3.7,3.7) {$d_1$};
%\draw[thick,dashed,arrows=->] (9.2,3.2) -- (8,4);
%\node at (9.2,3.7) {$d_{N-1}$};

%Placement
%\draw[thick,dashed] (3.8,7) edge[out=190,in=180,->] (0,1.5);
%\node at (0, 5.5) {Placement};

%Adversary?

%A file and subfiles
\draw (7.5, 8) -- (7.5, 8.5) -- (11.5, 8.5) -- (11.5,8) --(7.5,8);
\node at (7.1, 8.2) {{\scriptsize $W_i$}};
\draw (8.3,8)--(8.3,8.5);
\node at (7.8,8.7) {{\tiny $W_{i,0}$}};
\draw (9.1,8)--(9.1,8.5);
\node at (8.8,8.7) {{\tiny $W_{i,1}$}};
\node at (10, 8.2) {$\cdots$};
\draw (10.7,8)--(10.7,8.5);
\node at (11,8.7) {{\tiny $W_{i,n-1}$}};
\end{tikzpicture}
  
  \caption{A network model for coded caching}
  \label{fig:caching}
\end{figure}
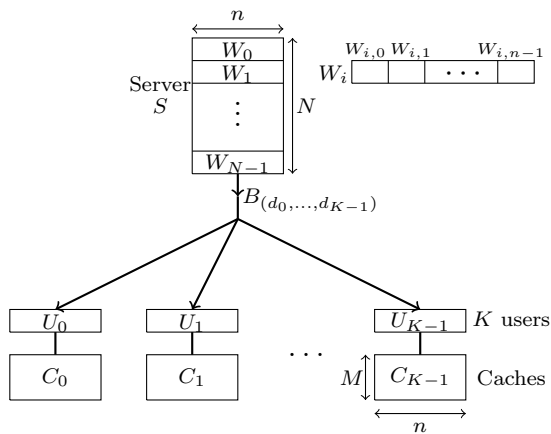

The  model for coded caching established by Maddah-Ali and Niesen in \cite{MaddahAli2014} forms the basis for those considered in most later research in the area.  We describe it as follows (see Figure~\ref{fig:caching}). 

 A central server $S$ stores $N$ files, $W_0, \ldots, W_{N-1}$, each of
 size $n$ units over some alphabet $\Sigma$.
 The server connects via a single  (error-free) link to $K$ users $\{U_0, U_1, \ldots, U_{K-1}\}$.  User $U_i$ has
 a local cache $C_i$ of size $Mn$, i.e., capable of storing an amount of data equivalent to $M$ files.
We write $W_i = (W_{i,0},
W_{i, 1}, \ldots, W_{i, n-1})$.

In the \textit{placement phase} the user caches are filled.
We refer to the content of user $U_i$'s cache as $C_i$.

Each user then \textit{demands} one of the $N$ files
stored by the central server.\footnote{We note that this is not treated as a ``phase'' in the literature - normally there are only two phases described: the placement phase and the 
delivery phase. For clarity, and with security in mind, we make this phase explicit here.}  We express the list of demands as $\bm{d} = (d_0, \ldots, d_{K-1})$ to indicate that user $U_i$ demands file $W_{d_i}$. 

In the \textit{delivery phase} the server broadcasts some message
$B_{\bm{d}}$ so that each user's demand is satisfied.  

We will refer to this as a centralised scheme. There are
variations to the network model, which we mention later,
but our primary focus is the security requirements and properties of this fundamental model, along with their
interactions with each other.

%This is the model of \cite{MaddahAli2014} and seems to be the accepted
%model in the literature. 

The aim of coded caching is to reduce the amount of data transferred over the shared link during the delivery phase.  This is expressed by a parameter $R$ that is usually referred to in the coded caching literature as the ``rate'' of the scheme, which is defined as the expected amount of data transfer divided by the size of a file. Thus in designing a coded caching scheme we are looking to minimise the rate.\footnote{This use of the word {\em rate} is in contrast with its use in other contexts, such as in coding theory, 
where it typically denotes a value between 0 and 1 that one seeks to maximise. Some later literature uses the perhaps more apt term {\em load} to refer to the same quantity $R$.} We note that another parameter involved in measuring the efficiency of coded caching is subpacketisation - the number of parts $n$ into which each file is divided.   Some schemes that focus only on improving the rate do involve impractical levels of subpacketisation. We mention this when relevant, in our discussion of \textit{secure} coded caching in Section \ref{sec:survey}.

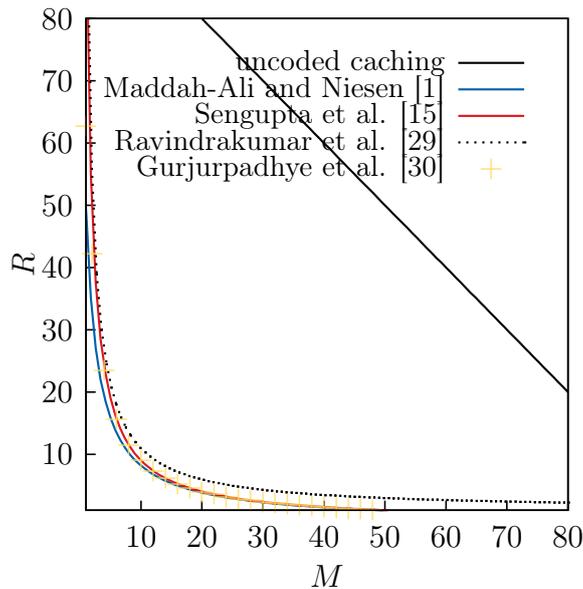
\begin{figure}[htb]
\centering
%\include{ratememorycurves}

% GNUPLOT: LaTeX2e picture (pict2e)
\setlength{\unitlength}{0.120450pt}
\ifx\plotpoint\undefined\newsavebox{\plotpoint}\fi
\ifx\transparent\undefined%
    \providecommand{\gpopaque}{}%
    \providecommand{\gptransparent}[2]{\color{.!#2}}%
\else%
    \providecommand{\gpopaque}{\transparent{1.0}}%
    \providecommand{\gptransparent}[2]{\transparent{#1}}%
\fi%
\begin{picture}(1889,1889)(0,0)
\miterjoin\buttcap
\color{black}
\sbox{\plotpoint}{\rule[-0.300pt]{0.600pt}{0.600pt}}%
\linethickness{0.6pt}%
\Line(267,440)(297,440)
\put(226,440){\makebox(0,0)[r]{$10$}}
\Line(267,635)(297,635)
\put(226,635){\makebox(0,0)[r]{$20$}}
\Line(267,830)(297,830)
\put(226,830){\makebox(0,0)[r]{$30$}}
\Line(267,1025)(297,1025)
\put(226,1025){\makebox(0,0)[r]{$40$}}
\Line(267,1220)(297,1220)
\put(226,1220){\makebox(0,0)[r]{$50$}}
\Line(267,1415)(297,1415)
\put(226,1415){\makebox(0,0)[r]{$60$}}
\Line(267,1610)(297,1610)
\put(226,1610){\makebox(0,0)[r]{$70$}}
\Line(267,1805)(297,1805)
\put(226,1805){\makebox(0,0)[r]{$80$}}
\Line(438,265)(438,295)
\put(438,182){\makebox(0,0){$10$}}
\Line(627,265)(627,295)
\put(627,182){\makebox(0,0){$20$}}
\Line(817,265)(817,295)
\put(817,182){\makebox(0,0){$30$}}
\Line(1007,265)(1007,295)
\put(1007,182){\makebox(0,0){$40$}}
\Line(1196,265)(1196,295)
\put(1196,182){\makebox(0,0){$50$}}
\Line(1386,265)(1386,295)
\put(1386,182){\makebox(0,0){$60$}}
\Line(1575,265)(1575,295)
\put(1575,182){\makebox(0,0){$70$}}
\Line(1765,265)(1765,295)
\put(1765,182){\makebox(0,0){$80$}}
\polygon(267,1805)(267,265)(1765,265)(1765,1805)
\put(1383,1666){\makebox(0,0)[r]{\small uncoded caching}}
\color[rgb]{0.00,0.00,0.00}
\sbox{\plotpoint}{\rule[-0.400pt]{0.800pt}{0.800pt}}%
\linethickness{0.8pt}%
\Line(1424,1666)(1629,1666)
\polyline(627,1805)(630,1802)(645,1786)(660,1771)(676,1755)(691,1740)(706,1724)(721,1709)(736,1693)(751,1678)(766,1662)(781,1646)(797,1631)(812,1615)(827,1600)(842,1584)(857,1569)(872,1553)(887,1538)(903,1522)(918,1506)(933,1491)(948,1475)(963,1460)(978,1444)(993,1429)(1008,1413)(1024,1398)(1039,1382)(1054,1366)(1069,1351)(1084,1335)(1099,1320)(1114,1304)(1129,1289)(1145,1273)(1160,1258)(1175,1242)(1190,1226)(1205,1211)(1220,1195)(1235,1180)(1251,1164)(1266,1149)(1281,1133)(1296,1118)(1311,1102)(1326,1086)(1341,1071)(1356,1055)(1372,1040)(1387,1024)(1402,1009)(1417,993)(1432,978)(1447,962)(1462,946)(1478,931)(1493,915)(1508,900)(1523,884)(1538,869)(1553,853)(1568,838)(1583,822)(1599,806)(1614,791)(1629,775)(1644,760)(1659,744)(1674,729)(1689,713)(1704,698)(1720,682)(1735,666)(1750,651)(1765,635)
\color{black}
\put(1383,1583){\makebox(0,0)[r]{\small Maddah-Ali and Niesen \cite{MaddahAli2014}}}
\color[rgb]{0.00,0.38,0.68}
\Line(1424,1583)(1629,1583)
\polyline(267,1210)(267,1210)(282,930)(297,774)(312,674)(328,605)(343,555)(358,516)(373,486)(388,461)(403,440)(418,423)(433,409)(449,396)(464,385)(479,375)(494,367)(509,359)(524,353)(539,346)(554,341)(570,336)(585,331)(600,327)(615,323)(630,319)(645,316)(660,313)(676,310)(691,307)(706,304)(721,302)(736,300)(751,298)(766,296)(781,294)(797,292)(812,290)(827,288)(842,287)(857,285)(872,284)(887,283)(903,281)(918,280)(933,279)(948,278)(963,277)(978,276)(993,275)(1008,274)(1024,273)(1039,272)(1054,271)(1069,270)(1084,270)(1099,269)(1114,268)(1129,267)(1145,267)(1160,266)(1175,265)
\Line(1175,265)(1190,265)
\color{black}
\put(1383,1500){\makebox(0,0)[r]{\small Sengupta et al. \cite{Sengupta2015}}}
\color[rgb]{0.87,0.09,0.12}
\Line(1424,1500)(1629,1500)
\polyline(273,1805)(282,1316)(297,980)(312,802)(328,692)(343,617)(358,563)(373,522)(388,490)(403,465)(418,443)(433,426)(449,410)(464,398)(479,386)(494,376)(509,368)(524,360)(539,353)(554,347)(570,341)(585,336)(600,331)(615,327)(630,323)(645,319)(660,316)(676,313)(691,310)(706,307)(721,304)(736,302)(751,299)(766,297)(781,295)(797,293)(812,292)(827,290)(842,288)(857,287)(872,285)(887,284)(903,283)(918,281)(933,280)(948,279)(963,278)(978,277)(993,276)(1008,275)(1024,274)(1039,273)(1054,272)(1069,271)(1084,270)(1099,269)(1114,269)(1129,268)(1145,267)(1160,267)(1175,266)(1190,265)
\Line(1190,265)(1205,265)
\color{black}
\put(1383,1417){\makebox(0,0)[r]{\small Ravindrakumar et al. \cite{Ravindrakumar2016}}}
\color[rgb]{0.40,0.40,0.00}
\multiput(1424,1417)(24.907,0.000){9}{\usebox{\plotpoint}}
\put(1629,1417){\usebox{\plotpoint}}
\multiput(273,1805)(0.476,-24.902){19}{\usebox{\plotpoint}}
\put(282,1334){\usebox{\plotpoint}}
\multiput(282,1334)(1.131,-24.881){14}{\usebox{\plotpoint}}
\put(297,1004){\usebox{\plotpoint}}
\multiput(297,1004)(2.139,-24.815){8}{\usebox{\plotpoint}}
\put(312,830){\usebox{\plotpoint}}
\multiput(312,830)(3.650,-24.638){5}{\usebox{\plotpoint}}
\put(328,722){\usebox{\plotpoint}}
\multiput(328,722)(5.013,-24.397){3}{\usebox{\plotpoint}}
\put(343,649){\usebox{\plotpoint}}
\multiput(343,649)(6.783,-23.965){3}{\usebox{\plotpoint}}
\put(358,596){\usebox{\plotpoint}}
\multiput(358,596)(8.557,-23.390){2}{\usebox{\plotpoint}}
\put(373,555){\usebox{\plotpoint}}
\multiput(373,555)(10.848,-22.420){2}{\usebox{\plotpoint}}
\put(388,524){\usebox{\plotpoint}}
\multiput(388,524)(12.814,-21.357){2}{\usebox{\plotpoint}}
\put(403,499){\usebox{\plotpoint}}
\multiput(403,499)(14.477,-20.267){2}{\usebox{\plotpoint}}
\put(418,478){\usebox{\plotpoint}}
\put(418.00,478.00){\usebox{\plotpoint}}
\put(433,460){\usebox{\plotpoint}}
\put(433.00,460.00){\usebox{\plotpoint}}
\put(449,446){\usebox{\plotpoint}}
\put(449.00,446.00){\usebox{\plotpoint}}
\put(464,433){\usebox{\plotpoint}}
\put(464.00,433.00){\usebox{\plotpoint}}
\put(479,422){\usebox{\plotpoint}}
\put(479.00,422.00){\usebox{\plotpoint}}
\put(494,412){\usebox{\plotpoint}}
\put(494.00,412.00){\usebox{\plotpoint}}
\put(509,404){\usebox{\plotpoint}}
\put(509.00,404.00){\usebox{\plotpoint}}
\put(524,396){\usebox{\plotpoint}}
\put(524.00,396.00){\usebox{\plotpoint}}
\put(539,389){\usebox{\plotpoint}}
\put(539.00,389.00){\usebox{\plotpoint}}
\put(554,383){\usebox{\plotpoint}}
\put(554.00,383.00){\usebox{\plotpoint}}
\put(570,378){\usebox{\plotpoint}}
\put(570.00,378.00){\usebox{\plotpoint}}
\put(585,372){\usebox{\plotpoint}}
\put(585.00,372.00){\usebox{\plotpoint}}
\put(600,368){\usebox{\plotpoint}}
\put(600.00,368.00){\usebox{\plotpoint}}
\put(615,364){\usebox{\plotpoint}}
\put(615.00,364.00){\usebox{\plotpoint}}
\put(630,360){\usebox{\plotpoint}}
\put(630.00,360.00){\usebox{\plotpoint}}
\put(645,356){\usebox{\plotpoint}}
\put(645.00,356.00){\usebox{\plotpoint}}
\put(660,353){\usebox{\plotpoint}}
\put(660.00,353.00){\usebox{\plotpoint}}
\put(676,350){\usebox{\plotpoint}}
\put(676.00,350.00){\usebox{\plotpoint}}
\put(691,347){\usebox{\plotpoint}}
\put(691.00,347.00){\usebox{\plotpoint}}
\put(706,344){\usebox{\plotpoint}}
\put(706.00,344.00){\usebox{\plotpoint}}
\put(721,341){\usebox{\plotpoint}}
\put(721.00,341.00){\usebox{\plotpoint}}
\put(736,339){\usebox{\plotpoint}}
\put(736.00,339.00){\usebox{\plotpoint}}
\put(751,337){\usebox{\plotpoint}}
\put(751.00,337.00){\usebox{\plotpoint}}
\put(766,335){\usebox{\plotpoint}}
\put(766.00,335.00){\usebox{\plotpoint}}
\put(781,333){\usebox{\plotpoint}}
\put(781.00,333.00){\usebox{\plotpoint}}
\put(797,331){\usebox{\plotpoint}}
\put(797.00,331.00){\usebox{\plotpoint}}
\put(812,329){\usebox{\plotpoint}}
\put(812.00,329.00){\usebox{\plotpoint}}
\put(827,327){\usebox{\plotpoint}}
\put(827.00,327.00){\usebox{\plotpoint}}
\put(842,326){\usebox{\plotpoint}}
\put(842.00,326.00){\usebox{\plotpoint}}
\put(857,324){\usebox{\plotpoint}}
\put(857.00,324.00){\usebox{\plotpoint}}
\put(872,323){\usebox{\plotpoint}}
\put(872.00,323.00){\usebox{\plotpoint}}
\put(887,321){\usebox{\plotpoint}}
\put(887.00,321.00){\usebox{\plotpoint}}
\put(903,320){\usebox{\plotpoint}}
\put(903.00,320.00){\usebox{\plotpoint}}
\put(918,319){\usebox{\plotpoint}}
\put(918.00,319.00){\usebox{\plotpoint}}
\put(933,318){\usebox{\plotpoint}}
\put(933.00,318.00){\usebox{\plotpoint}}
\put(948,317){\usebox{\plotpoint}}
\put(948.00,317.00){\usebox{\plotpoint}}
\put(963,315){\usebox{\plotpoint}}
\put(963.00,315.00){\usebox{\plotpoint}}
\put(978,314){\usebox{\plotpoint}}
\put(978.00,314.00){\usebox{\plotpoint}}
\put(993,313){\usebox{\plotpoint}}
\put(993.00,313.00){\usebox{\plotpoint}}
\put(1008,312){\usebox{\plotpoint}}
\put(1008.00,312.00){\usebox{\plotpoint}}
\put(1024,312){\usebox{\plotpoint}}
\put(1024.00,312.00){\usebox{\plotpoint}}
\put(1039,311){\usebox{\plotpoint}}
\put(1039.00,311.00){\usebox{\plotpoint}}
\put(1054,310){\usebox{\plotpoint}}
\put(1054.00,310.00){\usebox{\plotpoint}}
\put(1069,309){\usebox{\plotpoint}}
\put(1069.00,309.00){\usebox{\plotpoint}}
\put(1084,308){\usebox{\plotpoint}}
\put(1084.00,308.00){\usebox{\plotpoint}}
\put(1099,307){\usebox{\plotpoint}}
\put(1099.00,307.00){\usebox{\plotpoint}}
\put(1114,307){\usebox{\plotpoint}}
\put(1114.00,307.00){\usebox{\plotpoint}}
\put(1129,306){\usebox{\plotpoint}}
\put(1129.00,306.00){\usebox{\plotpoint}}
\put(1145,305){\usebox{\plotpoint}}
\put(1145.00,305.00){\usebox{\plotpoint}}
\put(1160,305){\usebox{\plotpoint}}
\put(1160.00,305.00){\usebox{\plotpoint}}
\put(1175,304){\usebox{\plotpoint}}
\put(1175.00,304.00){\usebox{\plotpoint}}
\put(1190,303){\usebox{\plotpoint}}
\put(1190.00,303.00){\usebox{\plotpoint}}
\put(1205,303){\usebox{\plotpoint}}
\put(1205.00,303.00){\usebox{\plotpoint}}
\put(1220,302){\usebox{\plotpoint}}
\put(1220.00,302.00){\usebox{\plotpoint}}
\put(1235,302){\usebox{\plotpoint}}
\put(1235.00,302.00){\usebox{\plotpoint}}
\put(1251,301){\usebox{\plotpoint}}
\put(1251.00,301.00){\usebox{\plotpoint}}
\put(1266,300){\usebox{\plotpoint}}
\put(1266.00,300.00){\usebox{\plotpoint}}
\put(1281,300){\usebox{\plotpoint}}
\put(1281.00,300.00){\usebox{\plotpoint}}
\put(1296,299){\usebox{\plotpoint}}
\put(1296.00,299.00){\usebox{\plotpoint}}
\put(1311,299){\usebox{\plotpoint}}
\put(1311.00,299.00){\usebox{\plotpoint}}
\put(1326,298){\usebox{\plotpoint}}
\put(1326.00,298.00){\usebox{\plotpoint}}
\put(1341,298){\usebox{\plotpoint}}
\put(1341.00,298.00){\usebox{\plotpoint}}
\put(1356,298){\usebox{\plotpoint}}
\put(1356.00,298.00){\usebox{\plotpoint}}
\put(1372,297){\usebox{\plotpoint}}
\put(1372.00,297.00){\usebox{\plotpoint}}
\put(1387,297){\usebox{\plotpoint}}
\put(1387.00,297.00){\usebox{\plotpoint}}
\put(1402,296){\usebox{\plotpoint}}
\put(1402.00,296.00){\usebox{\plotpoint}}
\put(1417,296){\usebox{\plotpoint}}
\put(1417.00,296.00){\usebox{\plotpoint}}
\put(1432,295){\usebox{\plotpoint}}
\put(1432.00,295.00){\usebox{\plotpoint}}
\put(1447,295){\usebox{\plotpoint}}
\put(1447.00,295.00){\usebox{\plotpoint}}
\put(1462,295){\usebox{\plotpoint}}
\put(1462.00,295.00){\usebox{\plotpoint}}
\put(1478,294){\usebox{\plotpoint}}
\put(1478.00,294.00){\usebox{\plotpoint}}
\put(1493,294){\usebox{\plotpoint}}
\put(1493.00,294.00){\usebox{\plotpoint}}
\put(1508,294){\usebox{\plotpoint}}
\put(1508.00,294.00){\usebox{\plotpoint}}
\put(1523,293){\usebox{\plotpoint}}
\put(1523.00,293.00){\usebox{\plotpoint}}
\put(1538,293){\usebox{\plotpoint}}
\put(1538.00,293.00){\usebox{\plotpoint}}
\put(1553,293){\usebox{\plotpoint}}
\put(1553.00,293.00){\usebox{\plotpoint}}
\put(1568,292){\usebox{\plotpoint}}
\put(1568.00,292.00){\usebox{\plotpoint}}
\put(1583,292){\usebox{\plotpoint}}
\put(1583.00,292.00){\usebox{\plotpoint}}
\put(1599,292){\usebox{\plotpoint}}
\put(1599.00,292.00){\usebox{\plotpoint}}
\put(1614,291){\usebox{\plotpoint}}
\put(1614.00,291.00){\usebox{\plotpoint}}
\put(1629,291){\usebox{\plotpoint}}
\put(1629.00,291.00){\usebox{\plotpoint}}
\put(1644,291){\usebox{\plotpoint}}
\put(1644.00,291.00){\usebox{\plotpoint}}
\put(1659,290){\usebox{\plotpoint}}
\put(1659.00,290.00){\usebox{\plotpoint}}
\put(1674,290){\usebox{\plotpoint}}
\put(1674.00,290.00){\usebox{\plotpoint}}
\put(1689,290){\usebox{\plotpoint}}
\put(1689.00,290.00){\usebox{\plotpoint}}
\put(1704,290){\usebox{\plotpoint}}
\put(1704.00,290.00){\usebox{\plotpoint}}
\put(1720,289){\usebox{\plotpoint}}
\put(1720.00,289.00){\usebox{\plotpoint}}
\put(1735,289){\usebox{\plotpoint}}
\put(1735.00,289.00){\usebox{\plotpoint}}
\put(1750,289){\usebox{\plotpoint}}
\put(1750.00,289.00){\usebox{\plotpoint}}
\put(1765,289){\usebox{\plotpoint}}
\color{black}
\put(1383,1334){\makebox(0,0)[r]{\small Gurjurpadhye et al. \cite{Gurjarpadhye2021}}}
\color[rgb]{1.00,0.86,0.35}
\put(267,1469){\makebox(0,0){$+$}}
\put(286,1067){\makebox(0,0){$+$}}
\put(324,703){\makebox(0,0){$+$}}
\put(362,549){\makebox(0,0){$+$}}
\put(400,469){\makebox(0,0){$+$}}
\put(438,421){\makebox(0,0){$+$}}
\put(476,388){\makebox(0,0){$+$}}
\put(514,365){\makebox(0,0){$+$}}
\put(552,348){\makebox(0,0){$+$}}
\put(590,334){\makebox(0,0){$+$}}
\put(627,323){\makebox(0,0){$+$}}
\put(665,315){\makebox(0,0){$+$}}
\put(703,307){\makebox(0,0){$+$}}
\put(741,301){\makebox(0,0){$+$}}
\put(779,296){\makebox(0,0){$+$}}
\put(817,291){\makebox(0,0){$+$}}
\put(855,287){\makebox(0,0){$+$}}
\put(893,283){\makebox(0,0){$+$}}
\put(931,280){\makebox(0,0){$+$}}
\put(969,277){\makebox(0,0){$+$}}
\put(1007,275){\makebox(0,0){$+$}}
\put(1045,272){\makebox(0,0){$+$}}
\put(1083,270){\makebox(0,0){$+$}}
\put(1120,268){\makebox(0,0){$+$}}
\put(1158,267){\makebox(0,0){$+$}}
\put(1526,1334){\makebox(0,0){$+$}}
\color{black}
\sbox{\plotpoint}{\rule[-0.300pt]{0.600pt}{0.600pt}}%
\linethickness{0.6pt}%
\polygon(267,1805)(267,265)(1765,265)(1765,1805)
\put(72,1035){\rotatebox{-270}{\makebox(0,0){$R$}}}
\put(1016,58){\makebox(0,0){$M$}}
\end{picture}

\caption{Rate/memory curves for coded caching schemes with $N=K=100$. Whereas uncoded caching leads to a rate that declines linearly with increasing memory, Maddah-Ali and Niesen's coded caching scheme gives a steep drop in rate for even a moderate increase in memory.  The Sengupta et al.\ scheme that provides content confidentiality, the scheme with file privacy of Ravindrakumar et al.\ and the demand private scheme of Gurjapadhye et al.\ also display a similar initial drop in rate.}
\label{fig:ratememory}
\end{figure}

In the absence of caching, delivering the $K$ files requested by the users over the link results in a rate of $K$.  Where uncoded caching is used, this can be reduced to 
\begin{equation}
K\left(1-\frac{M}{N}\right), 
\end{equation}
since for any user there is a probability of $M/N$ that the file they demand is already stored in their cache.  The factor $(1-M/N)$ is referred to as the {\em local caching gain}.  Maddah-Ali and Niesen's contribution was to provide a coded-caching scheme that attains rate 
\begin{equation}
K\left(1-\frac{M}{N}\right)\frac{1}{1+K\frac{M}{N}}, \label{eq:MArate}
\end{equation}
where the term $(1+KM/N)^{-1}$ is termed the {\em global caching gain}, and represents the additional improvement in rate that can be attributed to using coded as opposed to uncoded caching.  In general, the rate of a scheme is a function of the cache size $M$, and this can be visualised by plotting a {\em rate/memory curve} for a scheme.  Figure~\ref{fig:ratememory} depicts the rate/memory curves for various schemes in the case where $N=K=100$.  In particular, we can see that as $M$ increases, the global caching gain of the Maddah-Ali and Niesen scheme soon results in much lower rates than those obtained by the use of uncoded caching.  { Quantitative data underpinning these curves can be found in Table~\ref{tab:quantitative} in the Appendix.}

Maddah-Ali and Niesen show that the rate of their scheme is within a factor of 12 of the optimum attainable rate, which they denote $R^\ast(M)$ \cite{MaddahAli2014}.  In the literature, schemes whose rate is within a constant factor of an established bound are typically referred to as being {\em order optimal}.

\subsection{Assumptions in the model and variations}\label{subsec:assumptionsvariations}
%Assumptions in \cite{MaddahAli2014}: 
Assumptions in this commonly accepted model can be grouped as follows.

\begin{description}
\item[Constraints on bandwidth and storage] Firstly, it is assumed that the placement phase takes place during off-peak time and there is no constraint on the bandwidth at such time.  The only constraint is the cache size.  Secondly, the delivery phase takes place during peak time and the main concern is to reduce the amount of broadcast information required to satisfy user demand. 

\item[Access to information] It is assumed that placement takes place before user demand is known. In particular,  a user can demand any of the $N$ files, and a user may have any portion of any of the $N$ files during placement subject to cache size. It is assumed that this communication channel is error-free. User caches are assumed to be isolated - that is, they do not communicate with any entity apart from the user to whom they belong, and a user has no access to other users' caches. Furthermore, it is assumed that during the delivery phase only the server has access to the database.
Finally, the communication channel between the users and the server during delivery is a shared, error-free channel.
\end{description}

Other models have been studied in the literature; most of these consist of variations of the underlying model to accommodate scenarios with a more complicated channel between the server(s) and users. %where the variations stem from either the assumptions in the phases, or in the relationships between network entities.
\begin{description}%[\setlabelwidth{User Caches relationships}\usemathlabelsep]
\item[Placement phase] It seems to be an implicit assumption in much of the literature that cache placement is
a one-time event, and the cache content is unchanged over many demand-delivery phases. 
An exception is in the case of \textit{online coded caching} (introduced by Pedarsani, Maddah-Ali and Niesen in \cite{Pedarsani2016}),   
where user caches are updated as a function of the current cache content and the delivery of new content. Furthermore, it is commonly assumed that the server knows the content of user caches \cite{Birk1998}.  In the case of \textit{decentralised caching} \cite{MaddahAli2015}, cache placement is performed independently for each user and at delivery time the server has to be informed on active users and their cache content.  In any case, cache content is known at the time of delivery.  An exception to this is also described in \cite{MaddahAli2015}, where the server does not know the cache content and so has to send enough random linear combination of parts of the content to allow all demands to be satisfied.

\item[Channel properties] Various works remove the assumption that the channel broadcast is error-free, and instead model the channel as some specific type of lossy wireless channel, e.g., \cite{Shariatpanahi2017}. 

\item[Network topology] Generally it is assumed that there is only one server.  However, as mentioned above, \cite{Shariatpanahi2016,Shariatpanahi2017,Lampiris2018} allow multiple servers with various types of connections between the servers and the users.  
Karamchandani et al.\ \cite{Karamchandani2016} allow hierarchical levels - servers may populate mirrors which serve users.  Mirrors may communicate with each other.   On the other hand, there are proposals also for networks without online servers - \textit{device-to-device (D2D) coded caching} \cite{Zewail2016}: once the placement phase is over, user demands are satisfied by the network of users. 

\item[{User-Cache relationships}] It is generally assumed that a user has no knowledge of caches it has no access to.  
In works on shared caches, such as \cite{Wan2022,KrishnanNamboodiri2022}, a user has access to multiple caches.  In \cite{Peter2022}, multiple users share caches, with each user having access to only one cache.  In all cases the user has knowledge of caches assigned to it but has no knowledge of the content of other caches.

\end{description}

\subsection{Modelling security in coded caching}\label{subsec:securitymodelling}
A distinctive variation that can be added to the underlying model is that of considering \emph{secure} versions of coded caching. In the rest of this survey we restrict our attention to the simplest original model of  Maddah-Ali and Niesen \cite{MaddahAli2014}, illustrated in Figure \ref{fig:caching}, for, as we shall see, there are many issues still to be addressed in this fundamental scenario before these concepts can be consistently applied to more general channels or network topologies.

While there is general agreement on the aims of coded caching,
examination of the literature on secure coded caching reveals
two key issues.  Firstly, there is a lack of system-wide analysis.
Most proposals focus only on the security of individual parts of the process.  For
example, \cite{Gurjarpadhye2021} considers the privacy of
user requests (a user should not know other users' requests) but does
not address the confidentiality of the content - an eavesdropper may
be able to obtain parts of the content they are not entitled to.  By
contrast, \cite{Sengupta2015} addresses the
confidentiality of the content but not the privacy of user demands.
Meanwhile, other guarantees such as authenticity of demands and
integrity of broadcast messages are not addressed.  This narrow focus
is perhaps surprising given that the essential insight behind coded
caching is the improvements that can be realised from considering the
system as an integrated whole.

Secondly, different proposals consider different
network models under different assumptions, making it difficult to
compare or combine solutions in secure ways.  Crucial questions
for which consensus has not been established include the security
properties of the channels for the different phases, and the
capabilities of adversaries.  Establishing a comprehensive framework  
in which security properties of coded caching schemes can be analysed and discussed is key both to providing a meaningful comparison between existing proposals, and to establishing a rigorous theoretical understanding of the performance and security that such schemes are capable of delivering. 

In a step towards this, we survey existing proposals (Section~\ref{subsec:approaches}) in terms of both the models and assumptions they are working with as well as their security goals and the techniques they employ.  In Section~\ref{subsec:analysisanddiscussion} we summate our findings, identify security aspects that have not been adequately addressed by the literature to date, and highlight limitations of some commonly suggested techniques.

\section{On achieving secure coded caching}\label{sec:survey}

There exists a significant body of work that has addressed various security aspects of coded caching. From examining this, it is readily apparent that there is not a single definition for ``secure coded caching''.  Rather, there has been consideration of various security-related properties, either individually or in combination.  The general evolution of this work in the literature is that a specific security property is introduced, then there is a focus on determining optimal rates of coding caching schemes under that security model.  This typically involves a combination of information theoretic techniques to give lower bounds, as well as constructions of schemes secure in that model to give upper bounds on the optimal rate.  Some papers also look at constructing schemes that achieve good rates while restricting the amount of subpacketisation required.

In this section we describe the various security models that have been considered in the literature, in roughly chronological order of their introduction.  In each case we briefly summarise the contributions of papers working in the given model, and highlight the key techniques they use for constructing schemes that are secure in that model.  An overview is given in Table~\ref{tab:SCC}.

\subsection{Approaches in the literature}\label{subsec:approaches}
In this section we survey the literature on security aspects of coded caching, grouping papers according to the security properties they consider.  
%We restrict our attention to centralised caching scheme. 
\subsubsection{Content confidentiality}
While the general caching problem has been extensively studied in the literature, considerations of security only appeared with the work of Sengupta, Tandon and Clancy \cite{Sengupta2015} less than ten years ago. In their paper, the authors consider the problem of an \emph{eavesdropper} observing communication from the central server to the users over an insecure link with the aim of obtaining information on the contents of the files requested by the users; schemes that prevent this are said to have {\em secure delivery}.  Let $\operatorname{I}(X;Y)$ denote the mutual information of discrete random variables $X$ and $Y$ \cite{Welsh1988}.  Specifically, Sengupta et al. construct schemes in which $\operatorname{I}(W_0,\dotsc,W_{N-1};B_{(d_0,\dotsc,d_{K-1})})=0$. This property is also called {\em file secrecy} in the literature \cite{Qi2022}; we use the term {\em content confidentiality} to more clearly distinguish it from terminology used for other security properties.  The underlying approach to security is simple: let the server generate a set of secret keys, and place one in each cache in the placement phase. In the delivery phase, the server XORs the content it wishes to deliver to a cache with the relevant key, effectively using a one-time pad (OTP) to protect the content, which can be easily recovered by the cache as it stores the secret key.  

By careful subdivision of files and appropriate choice of key allocation, Sengupta et al.\ propose a scheme that attains rate 
\begin{equation}
K\left(1-\frac{M-1}{N-1}\right)\frac{1}{1+K\frac{M-1}{N-1}} \label{eq:sengupta}
\end{equation}
in the case $K\leq N$.  This rate is naturally higher than that required by schemes without security, as some of the cache space is occupied by keys rather than files; however, as $K$ and $N$ become large it asymptotically approaches the rate \eqref{eq:MArate}.  Figure~\ref{fig:ratememory} depicts the rate/memory curve for this scheme when $K=N=100$.  It is visibly close to that of that of the Maddah-Ali and Niesen scheme, particularly as $M$ grows larger (and thus the difference between $M$ and $M-1$ becomes less significant).  Thus, for practical values of $N$ and $K$, the overhead in the rate required for providing content confidentiality is minimal.
Sengupta et al.\ show that \eqref{eq:sengupta} is within a factor of 17 of the optimum attainable rate of any scheme delivering content confidentiality, which they denote $R^\ast_s(M)$.  Considered the seminal paper in the area,  \cite{Sengupta2015} has generated lots of follow-up work and extensions, including those discussed below. 

\subsubsection{File privacy}\label{sec:fileprivacy}
In 2016, Ravindrakumar, Panda, Karamchandani and Prabhakaran \cite{Ravindrakumar2016} introduce the notion of \emph{secretive}, as opposed to secure, coded caching where the objective is to ensure that no single user learns any information about files they did not request.  In a 2018 extended version of this work \cite{Ravindrakumar2018}, they change the nomenclature to {\em private} coded caching; this property is also referred to as {\em file privacy} in the literature \cite{Qi2022}. Let $\tilde{\mathbf{W}}_{d_i}$ denote the set of all files other than that demanded by user $i$.  A scheme has file privacy if $\operatorname{I}(\tilde{\mathbf{W}}_{d_i};B_{(d_0,\dotsc,d_{K-1})},C_i,d_i)=0$ for all $i$.  Although the security goal is different to that of content confidentiality, the same security techniques apply: they describe an optimal scheme for $N=K=2$ and $M=1$ where
the server generates and places independent and uniformly distributed secret keys in the caches during placement phase, and XORs these with the demanded file during delivery phase. A more general scheme where a threshold of keys is necessary is also proposed, achieved with secret sharing techniques. Interestingly, this scheme also attains content confidentiality. (Note that the optimal scheme for $N=K=2$ does not provide content confidentiality, as in cases where both users demand the same file the file may be transmitted in the clear.)

For $M=Nt/(K-t)+1$ with $t$ an integer between $0$ and $K-2$ inclusive, their scheme attains rate 
\begin{align}
&\phantom{=}\frac{K(N+M-1)}{N+(K+1)(M-1)}\nonumber\\&=K\left(1+\frac{M-1}{N}\right)\frac{1}{1+(K+1)\frac{M-1}{N}}. 
\end{align}
For general $M$ the convex hull of these points represents what is attainable.  This is shown to be within a constant factor of an information-theoretic lower bound on the rate \cite{Ravindrakumar2018}.  We see in Figure~\ref{fig:ratememory} that this rate is similar to \eqref{eq:sengupta} for small $M$, and slightly higher for larger $M$.  Note that a scheme without file privacy can attain rate 0 if the caches are large enough to store all the files (i.e.,\ if $M\geq N$).  However, this is not possible in the file privacy setting: as each user's cache contents reveal no information about any file, then arguments from information theory can be used to show that at least a file's worth of data must be transmitted in order for any user to recover their desired file.  As such, the rate of a scheme with file privacy cannot be less than 1.  Nonetheless, in practical terms the performance of this scheme is relatively close to order optimal schemes without file privacy. 

Meel and Rajan \cite{ShresthaPIMRC2021} discuss an approach to adding file privacy to coded caching schemes based on {\em placement delivery arrays}, a type of array that was introduced for designing coded caching schemes with low subpacketisation.  They show that the scheme of \cite{Ravindrakumar2018} can be viewed as a special case of this construction.  Extensions of this work to a setting with shared caches are given by the same authors in \cite{Shrestha2021}, and by Peter, Namboodiri and Rajan in \cite{PeterISIT2022}.  

Suthan, Chugh and Krishnan \cite{Suthan2017} show that the rate attained by \cite{Ravindrakumar2016} can be improved in a setting where you take into account common demands between users.
  
Zewail and Yener \cite{Zewail2016} address the security goals of \cite{Ravindrakumar2016} but in a setting of device-to-device (D2D), rather than centralised, communication. Here, the placement phase is as usual run by the central server, but the delivery phase is done D2D.  As in \cite{Ravindrakumar2016}, the aim is for each user to recover the file they demand, but not receive information about other files. The focus of the paper is on achievability and bounds, like most of the literature, so the security solution is not formalised or discussed in depth. However, the proposed approach is similar to \cite{Sengupta2015}, having the server distribute secret keys, with the addition of secret sharing (SS) techniques to address the D2D setting.

\subsubsection{Demand privacy}
Demand privacy refers to the property that the delivery phase should reveal no information to any user about which files the remaining users requested, so we have $\operatorname{I}(d_i;B_{(d_0,\dotsc,d_{K-1})},C_j,d_j)=0$ for all $i \neq j$.  This was considered in 2019 by Wan and Caire \cite{Wan2021}, although they work in a model where the server has the additional ability to perform multicast communications to chosen subsets of users.  Kamath \cite{kamath2019demand} observes that it is possible to construct a coded caching scheme with demand privacy for $N$ files and $K$ users from a non-private scheme with $N$ files and $NK$ users, an approach also considered in \cite{Wan2021}.  This is sometimes referred to as a {\em virtual users} approach; each user is associated with $N$ of the $NK$ virtual caches, and the fact that a user cannot tell which virtual cache a given user is using to recover their file is used to obscure which file they have requested.  It is mentioned in the papers \cite{Gurjarpadhye2021, Kamath2020} that demands are conveyed secretly from the user to the server, although no detail is provided regarding this phase - as we have highlighted, it is typically not formalised nor described. What is described is that, in the placement phase, the server shares secret keys with the user by placing them in the cache, and in the delivery phase these will be used as OTPs.  A series of papers \cite{Kamath2020, Gurjarpadhye20}, culminating in Gurjarpadhye, Ravi, Kamath, Dey and Karamchandani's \cite{Gurjarpadhye2021} use variations of this approach to construct a range of schemes that between them provide an order optimal rate for demand private schemes for all relevant combinations of parameters.  Figure~\ref{fig:ratememory} depicts the rate obtained when $K = N = 100$ (with the parameter $r$, related to sizes of segments of a file, set to $N-1$ = 99, giving the best rate)  This is visibly close to that obtained by the scheme of Sengupta et al.\ that provides content confidentiality \cite{Sengupta2015}.

This technique of carefully placed secret keys in caches to be used as OTP was also used in constructing demand-private schemes where users have access to multiple caches (\cite{Namboodiri2022privacy}).

An extension to this work is \cite{Aravind2020} by Aravind, Sarvepalli and Thangaraj, who focus on efficiency (in particular, on how to keep subpacketisations of files low) and discuss a privacy variant called \emph{partial privacy} in which a user should still not learn another user's demand, but may obtain partial information about the overall set of demands.  Their construction techniques include starting with non-private coded caching schemes that use placement delivery arrays to reduce subpacketisation and modifying them to obtain schemes with demand privacy.

Another scheme providing demand privacy and content confidentiality is given by Cheng, Liang and Wei \cite{Cheng2020}.  They base their scheme on a structure they introduce called a {\em secure placement delivery array} (SPDA), which is a variant of the previously-studied placement delivery arrays.

%\subsubsection{File and Demand Privacy}
Qi and Ravi \cite{Qi2022} consider schemes that provide both file privacy and demand privacy. They show that satisfying both properties implies the scheme also gives content confidentiality, whereas file privacy or demand privacy alone do not. Their results on the rate of such schemes focus on the particular cases of $N=2$ and $K=2$, or the minimal-memory parameters $M=1$ and $R=K$ and the minimal-rate parameters $M=N(K-1)+1$ and $R=1$. The techniques suggested in the paper include the use of the one-time pad, permutation functions and secret sharing.
%\textcolor{orange}{\subsubsection{Demand Privacy with Cache Privacy}

Finally, Gholami, Wan, Sun, Ji and Caire \cite{Gholami2022} consider schemes that provide demand privacy, and additionally prevent users from learning which combinations of file packets are stored in other user's caches (they refer to this as {\em private caches}).  Note that what is being kept private here is the metadata consisting of the identity of the packets/combinations of packets, rather than the values of the packets themselves.  They show that schemes with demand and cache privacy can be constructed from a 2-server private information retrieval (PIR) scheme, by essentially treating the data stored in the caches as the data transmitted by one of the servers in the PIR scheme.

\subsubsection{Colluding users}
Work by Ma, Shao and Shao considers security in t
he presence of up to $\ell$ \emph{colluding users} \cite{Ma2021}, who could potentially share their cache content.  This paper considers schemes that seek to provide privacy of files against sets of up to $\ell$ colluding users prior to the delivery phase, as well as privacy after the delivery phase of files not requested by colluding users.  Specifically, for any set $i_0,\dotsc,i_{\ell-1}$ of colluding users they require both $\operatorname{I}(W_0,\dotsc,W_{N-1};C_{i_0},\dotsc,C_{i_\ell})=0$, and $\operatorname{I}(\tilde{\mathbf{W}}_{d_{i_0},\dotsc,d_{i_{\ell-1}}};B_{(d_0,\dotsc,d_{K-1})},C_{i_0},\dotsc,C_{i_{\ell-1}})=0$.  No details of the formal analysis are provided with respect to the security guarantees promised, and practical issues such as key distribution and key updates are not addressed.  As in other coded caching schemes the rate initially drops quickly as $M$ increases, but then declines more gradually, reaching a positive value when $M=N$ that increases with the number of colluding users that the scheme can support.

Yan and Tuninetti \cite{Yan2021} propose schemes that do not rely on virtual users, and hence achieve a lower degree of subpacketisation.  These use techniques similar to those seen in constructing private information retrieval schemes.  They also consider schemes in which users can demand not just a single file, but also a linear combination of files.  Namboodiri and Rajan \cite{Namboodiri2021}
use MDS codes (in an approach directly related to a threshold secret sharing scheme \cite{mceliecesarwate}) to produce schemes that meet the lower bounds established by Yan and Tuninetti when the cache sizes are small.

\begin{table}[h!]
\renewcommand{\arraystretch}{1.3}
 {\centering
  \caption{Overview of Secure Coded Caching State of the Art} 
  \label{tab:SCC}
%\begin{tabular}{ p{1.5cm}|p{1.2cm}|p{2.7cm}|p{1.8cm}  }
\begin{tabular}{ rlp{4.5cm}l }
 %\hline
 %\multicolumn{4}{|c|}{Overview of Secure Coded Caching State of the Art} \\
 \hline
% & {Security } & {} & {Main}\\
%{Scheme} & {Goal} & {Distinctive Feature} & {Technique}\\
 {Scheme} & {Security goal} &  {Distinctive feature} & {Main technique}\\
 \hline
\cite{Sengupta2015}  & CC  & {seminal paper} &   OTP \\
\cite{Zewail2016}&   FP   & {D2D}   &OTP, SS\\
\cite{Ravindrakumar2016,Ravindrakumar2018} &FP  & {secretive/private} &  OTP, SS\\
\cite{ShresthaPIMRC2021} &FP & uses PDAs in schemes with FP  &  PDA\\
\cite{Shrestha2021,PeterISIT2022} &FP &  shared caches &  PDA\\
\cite{Wan2021} &DP & requires multicast &  VU\\
 \cite{kamath2019demand,Kamath2020}, \\ \cite{Gurjarpadhye20,Gurjarpadhye2021} & DP   & \raggedright{demands to server are secret} & VU,OTP\\
\cite{Namboodiri2022privacy} & DP &\raggedright{multi-access caches} & OTP \\ 
\cite{Aravind2020} & DP*   &\raggedright {partial privacy} & VU, PDA\\
\cite{Cheng2020} &   CC, DP  &\raggedright rate, subpacketisation independent of $N$ &SPDA\\
\cite{Qi2022}& DP, FP&\raggedright shows DP\&FP$\Rightarrow$CC&OTP, SS\\
\cite{Gholami2022}&DP, CP& cache privacy &PIR\\
\cite{Ma2021} & CC, DP, FP &\raggedright  {collusions, informal} &  OTP, SS \\
\cite{Yan2021}& DP*&\raggedright{collusions, linear combination demands}&PIR*\\
\cite{Namboodiri2021}&DP*&\raggedright{collusions, small caches}&MDS codes\\
\cite{Engelmann2017}& DP* & \raggedright{computational security, informal}& AES\\
 \hline
\multicolumn{4}{p{12cm}}{\emph{Notation:} AES: Advanced Encryption Standard, CC: Content Confidentiality, CP: cache privacy, DP: Demand Privacy, FP: file privacy,  OTP: one-time pad, PDA: placement delivery array, PIR: private information retrieval,  SPDA: secure placement delivery array, SS: secret sharing, VU: virtual users, * denotes a variant of a concept.}
\end{tabular}  }
\end{table}

\subsubsection{Subtle privacy concerns}
In a departure from the rest of the literature, Engelman and Elia \cite{Engelmann2017} consider subtle privacy concerns arising from coded caching, namely the lack of user privacy if association of a user demand is possible (akin to demand privacy), and unauthorised tracking of statistics of file requests themselves. 

The authors discuss the challenges in using https in this context, they refer to proposals for encryption in caches \cite{LPQS17} and identify sought-after privacy properties. These are discussed rather than formalised, and a solution is described where both the system model and used techniques are different from previous approaches. More specifically, \cite{Engelmann2017} separates the entity of the server into the content provider and internet service provider (ISP), wishing to minimize trust in the ISP, and suggests the use of phantom users and symmetric encryption (namely, AES) of files as techniques to achieve privacy.

Table~\ref{tab:SCC} contains an overview of the existing security-relevant literature categorised in terms of their security goals, distinctive features and techniques used in constructing secure schemes.  As we can see from Figure~\ref{fig:ratememory}, it is possible to design schemes providing content confidentiality, file privacy, or demand privacy that achieve rates comparable to the (insecure) Maddah-Ali and Niesen scheme for practical parameter sizes.  This is encouraging news, as it demonstrates that it is possible to introduce security to coded caching while still achieving the sort of reduction in rate that is the whole reason for considering coded caching.  However, it is apparent that the literature to date has been driven by a desire to establish optimal rates in these specific new models, rather than by a system-wide perspective on security as a whole.

In Section~\ref{subsec:analysisanddiscussion} we focus specifically on the assumptions and techniques that have been used to construct schemes in these various models, and identify limitations that are common to these approaches.

\subsection{Analysis of the state of the art: Common features and limitations from a security perspective}\label{subsec:analysisanddiscussion}
We observe that in addressing security in the context of coded caching, several assumptions have been made consistently throughout the literature discussed in Section~\ref{subsec:approaches}. Specifically:

\begin{enumerate}
\item The server storage is assumed to be secure.
\item It is assumed there is a secure and private channel between the server and each user during the \emph{placement phase}.
\item It is assumed there is a secure and private channel between each user and the server during the \emph{demand phase}.  Indeed, as we have discussed before, the {demand phase} is never modelled.
\item It is commonly assumed that the broadcast channel in the \emph{delivery phase} is integrity-protected but not secure against wiretappers.
\end{enumerate}

These arguably strong assumptions have led the research community to focus on only two specific types of security goals: 

\begin{description}
\item[Confidentiality properties] These include content confidentiality, where files are kept confidential from an external eavesdropper, and file privacy, where they are kept confidential from users who did not request them.

\item[Privacy properties] These include demand privacy, where the server's broadcast does not leak information about users' demands to other users, and cache privacy, where it does not leak information to other users about which files are stored in users' caches. 

\end{description}
In both cases the only type of adversary considered is honest-but-curious, and could be viewed in general as an external adversary who potentially has access to the contents of one or more users' caches.

The main goal of the research to date has been to investigate the rate/memory trade-off for various security models, and schemes have been designed with a view to give bounds on that trade-off (sometimes these incorporate other trade-offs, such as the desire to reduce subpacketisation).  Unsurprisingly, this has tended to involve the use of information-theoretic primitives previously studied in other contexts, such as the one-time pad and secret sharing techniques.

Both the models considered, and the techniques employed, have limitations that become apparent if you start from a viewpoint of considering security requirements for the system as a whole.
\begin{description}
\item[Limited range of security goals]

Features such as integrity or authentication of the communications during the placement, demand and delivery phases of the scheme are not considered.

\item[Limited adversarial models]

Schemes to date have only considered {\em  honest-but-curious} adversaries, and have not taken into account the stronger attacks possible for an {\em active} adversary who can modify values in the system. If we extend the model to consider multiple rounds of content delivery, it may also be desirable to consider adversaries who have access to additional information, such as which files were requested by users in previous rounds. Furthermore, one could also consider the setting in which the server itself is untrusted.

\item[Limited techniques]
%Practical issues lead to security issues

The use of the OTP (or indeed, more sophisticated information-theoretic techniques such as secret sharing) comes with various practical issues.  There is the well-known fact that an OTP requires at least as much entropy as the information we are protecting.  This is seen in \cite[Example 1]{Sengupta2015}, where the secret key stored in a user's cache has to be the same length as the broadcast from the server.  This aspect has been a main consideration of such papers, as the need to devote cache space to keys reduces the amount available for file storage and hence affects the rate/memory trade-off.  { While OTPs have the advantage of low latency in that it is very efficient at the operation end, }  there are additional important practical issues that have not received the same attention.  For instance, proposed solutions do not detail how the secret keys are placed in the cache, and do not consider the practical need to update such keys after every use.  The following example illustrates this using the achievable scheme of \cite[Theorem 1]{Sengupta2015}.

\end{description}

\begin{example}

An OTP, as the name implies, is designed to be used only once.  For example, consider the case in  \cite[Example 1]{Sengupta2015} where there are two users, $U_1$ with cache $C_1$, and $U_2$ with cache $C_2$. User $U_i$, $i=1, 2$, is given secret key $K_i$ to store in $C_i$ during the placement phase. The files are $W_1=A$, $W_2=B$, and for demand $(d_1, d_2)$, the server broadcasts $(W_{d_1}\oplus K_1, W_{d_2}\oplus K_2)$.  If the secret keys are not renewed, an eavesdropper will be able to tell if a user demanded the same file twice.  If any part of a file is known the eavesdropper will also be able to deduce some values of the secret key.  If both users demand the same file they will be able to discover the other user's secret key.  One can imagine the scenario of, say, a content streaming service, where placement was performed in the day when demand is not high, and many rounds of demands and deliveries performed in the evening.  Thus information can be leaked to an eavesdropper.

Another limitation of an OTP is that it provides no defence against an active adversary who can modify communications.  For example, any modification of $W_{d_1}\oplus K_1$ will not be detected and user $U_1$ can potentially be provided with corrupted information.  Not only does this prevent the scheme from attaining the desired functionality, it also has the potential to be used as part of an attack on other properties such as privacy or confidentiality.  
\end{example}

In essence, while it is the first attempt to add security to the process, the literature so far has taken an oversimplified approach which in practice would not satisfy modern applications. 

{ Indeed, we explore next what the most common application scenarios would require in terms of security, and in Section \ref{Open} we discuss in more detail these issues highlighting relevant research directions.}

\begin{table}[!t]
	\centering
	\renewcommand{\arraystretch}{1.3}
	\scriptsize
	\begin{tabular}{lccccccc}
		
		\textbf{Application} & 
		\rot{Passive\\attacker} & 
		\rot{Active\\attacker} & 
		\rot{Content\\confidentiality} & 
		\rot{File\\privacy} & 
		\rot{Demand\\privacy} & 
		\rot{File\\integrity} & 
		\rot{Server\\authentication} \\
		\hline\hline
		Video streaming   & \fullcircle & \emptycircle & \emptycircle & \emptycircle & \fullcircle & \emptycircle & \emptycircle \\
		Software updates  & \fullcircle & \fullcircle  & \emptycircle & \emptycircle & \emptycircle & \fullcircle & \fullcircle \\
		Online gaming     & \fullcircle & \emptycircle & \emptycircle & \emptycircle & \emptycircle & \fullcircle & \fullcircle \\
		Cloud             & \fullcircle & \fullcircle  & \fullcircle  & \fullcircle  & \fullcircle  & \fullcircle & \fullcircle \\
		\hline\hline
	\end{tabular}
	\caption{{A mapping of security requirements for different SCC applications.}}
	\label{tableapp}
\end{table}

\subsection{{Applications of SCC and their security requirements}}

{To explore and contextualise the security properties we have discussed so far, we consider four different CDN uses and establish their  security requirements, highlighting where some are more relevant than others. Our discussion is summarised in Table \ref{tableapp}.}

{\paragraph{Video Streaming} 
The primary concern in video streaming is protecting user privacy (which movies or shows are watched) from passive observers, since viewing patterns can reveal personal interests or habits. Content confidentiality is less important because the material is already widely available, often governed separately by DRM. File integrity and server authentication are not critical, as tampering would be obvious and providers are well-known.}

{\paragraph{Software Updates} 
In software updates it is paramount these come from authentic servers and remain unmodified, otherwise they could deliver malware. Integrity, server authentication, and protection against active attackers are therefore essential. Confidentiality and file privacy are less critical, as updates are public once released.}

{\paragraph{Online Gaming} 
In the context of online gaming, integrity is crucial to stop tampered assets or cheats, and server authentication ensures players connect to legitimate servers rather than malicious copies. Content confidentiality, demand privacy, and file privacy are less relevant because game data is not secret in the same way as personal documents.}

{\paragraph{Cloud Services} 
Cloud storage and processing often involve highly sensitive files (business data, personal photos, health records). Hence, all security dimensions apply: confidentiality, privacy of stored content, demand privacy (access patterns), file integrity, and robust server authentication. Both passive and active attacks must be resisted to maintain trust in cloud providers.}\\

{This overview of applications and requirements contextualises the need for security provided by SCC.}

\section{Secure coded caching and related primitives}\label{sec:related}

\label{sec:relatedprimitives}

Given that formal and practical security considerations have not been at the forefront of the research in secure coded caching (SCC), as we have discussed in Section \ref{sec:survey}, 
we are motivated to explore whether this area could benefit from results in other fields, where the path to a more comprehensive formalisation has been further developed.
Therefore in this section we consider SCC in the context of related primitives. In particular, we examine its relationship to network coding, index coding, private information retrieval, and broadcast encryption. These primitives are selected both for their functional affinity, i.e., their focus on content delivery/retrieval, and for the maturity of their security frameworks, featuring well-established security definitions and extensively studied solution paradigms. This makes them a useful point of comparison.
For each primitive we give a brief description and then we discuss how SCC compares mainly in terms of functionality and  security provided. Our aim is to understand whether it is possible to capitalise upon the rich existing literature and get insights into improving the state of the art in SCC.

\subsection{Secure network coding}
\label{sub:SNC}

In the seminal paper \cite{Ahlswede2000}, Ahlswede, Cai, Li and Yeung propose network coding as a solution to the
situation where there is one source emitting $N$ messages that travel via intermediate nodes
to multiple receivers, each wanting a subset of the $N$ messages.
The network is modelled as a directed acyclic graph: the source, receivers, and intermediate 
nodes are vertices of the graph, while an edge indicates an error-free link between
two nodes.  A network code
is a set of functions, one associated with each edge, and each
function takes the inputs to a node (i.e., data packets), combines them in some way, and outputs
some data that
becomes the input to the node that is head of the edge. The
combining of data packets is aimed at optimising the amount
of information that can be passed through a network where the
links have limited bandwidth, and this works if each receiver
is be able to reconstruct the intended subset of messages from
what it receives.

\begin{figure}[htb]
  \begin{center}
\begin{tikzpicture}[scale=0.60]
%\draw[step=1cm,gray,very thin] (0,0) grid (10,13);

%The users
\draw (2,1) circle (0.5);
\node at (2,1) {{\scriptsize $U_0$}};
\draw (8,1) circle (0.5);
\node at (8,1) {{\scriptsize $U_1$}};

%Intermediate nodes
\draw (1,4) circle (0.5);
\node at (1,4) {{\scriptsize $C_0$}};
\draw (5,4) circle (0.5);
\node at (5,4) {{\scriptsize $E_S$}};
\draw (9,4) circle (0.5);
\node at (9,4) {{\scriptsize $C_1$}};

%The server
\draw (5,7) circle (0.5);
\node at (5,7) {{\scriptsize $S$}};

%Source to intermediate nodes
\draw (4.5,7) edge[out=190,in=80,->] (1,4.5);
\draw[arrows=->] (5,6.5) -- (5,4.5);
\draw (5.5,7) edge[out=350,in=100,->] (9,4.5);
\node at (1.5,6.6) {{\scriptsize $W_0, \ldots, W_{N-1}$}};
\node at (6.6,5.5) {{\scriptsize $W_0, \ldots, W_{N-1}$}};
\node at (8,7) {{\scriptsize $W_0, \ldots, W_{N-1}$}};

%Intermediate nodes to sink
\draw (1,3.5) edge[out=270,in=120,->] (1.9,1.5);
\draw[arrows=->] (4.5,3.8) -- (2.1,1.5);
\draw[arrows=->] (5.5,3.8) -- (7.9,1.5);
\draw (9,3.5) edge[out=270,in=60,->] (8.1,1.5);
\node at (1,2.4) {{\scriptsize $c_0$}};
\node at (9.1,2.4) {{\scriptsize $c_1$}};
\node at (3,3.5) {{\scriptsize $B_{(d_0, d_1)}$}};
\node at (6.0,2.4) {{\scriptsize $B_{(d_0, d_1)}$}};
\node at (2,0.2) {{\scriptsize $W_{d_0}$}};
\node at (8,0.2) {{\scriptsize $W_{d_1}$}};
\end{tikzpicture}
  \end{center}
  \caption{Coded caching as network coding}
  \label{fig:SNC}
\end{figure}
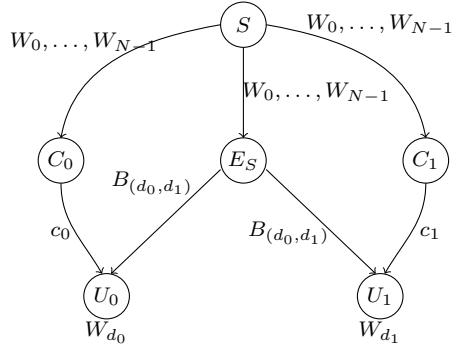

In \textit{secure} network coding, there are two main concerns in general, both to do with adversaries who have
access to a set of edges.  In the first instance (e.g., \cite{Cai2011, Liu2015, Wang2019}), the concern is that adversaries who can \emph{eavesdrop}
on the information passed along these edges should not gain any knowledge of what the source 
has transmitted.  

The second instance involves Byzantine adversaries who are capable of adding or modifying
communications at edges (as in \cite{Ho2004, Charles2006, Zhao2007}). 

In common with coded caching, network coding aims at delivering the right set
of messages to receivers and uses coding to optimize network performance.  One could model coded caching as network coding using
directed acyclic graphs, but the graph would be quite restricted: the only intermediate nodes
are the user caches.
We illustrate how this might work in the following example (See also Figure \ref{fig:SNC}).

\begin{example}\label{eg:networkcoding}

Consider a coded caching example where a central server $S$ stores $N$ files $W_0, W_1, \ldots, W_{N-1}$,
each of the same size $n$ over some alphabet.  The server connects via a broadcast link to two users
$U_0$, $U_1$.   User $U_i$, $i=0,1$, has a local cache $C_i$ of size $Mn$.  

In the placement phase the user caches are filled: the content $c_i$ is a function of $\{W_0, W_1, \ldots, W_{N-1}\}$ - we will denote them $c_0$ and $c_1$, as outputs of the functions. Later $U_0$ demands $W_{d_0}$, and $U_1$ demands $W_{d_1}$.  The server $S$ encodes the broadcast message $B_{(d_0, d_1)}$ using some encoding function $E_S$ so that both users' demands can be fulfilled.

We can model this as network coding: the server is the source, the users are the sinks, and the broadcast and caches are intermediate nodes.  The functions $E_S$, $c_0$, $c_1$ are fixed.    The edges have different capacity: $S \to C_0$,  $S \to C_1$, $S \to E_S$ have capacity $Nn$,  $C_0 \to U_0$, $C_1 \to U_1$ have capacity $Mn$ (See Figure \ref{fig:SNC} ). The flow $E_S \to U_0 $ (or $ E_s \to U_1$) is the amount of data to be broadcast to satisfy user demands, and is what coded caching aims to minimise.
\end{example}

While functionality-wise the above example suggests a potential translation between the two primitives, we note that they approach network modelling quite differently: in secure network coding all links have the same security properties, while in secure coded caching the user-cache links, the server-cache links, and the server-user link may have different properties and trust assumptions. 
In addition, the standard model of network coding does not include a placement phase. In our modelling of secure coded caching as network coding we capture placement as the function associated with the edge between the source and the caches, as described in Example \ref{eg:networkcoding}. 

From a security standpoint, secure network coding aims to protect against eavesdroppers as well as active adversaries, while SCC has only considered eavesdroppers, and only in specific phases.
Therefore intuitively there would be some security gains from the translation, but, concretely, network coding's lack of user demand modelling and the ability to capture only one instance of the coded caching placement-delivery cycle would result in a very restrictive SCC construction.

On the other hand, attempting to base secure network
coding on SCC does not seem promising.
Firstly, we note that the simple topology of SCC's network, in particular the lack of multiple intermediate nodes, hinders the maximisation in network throughput network coding is designed for. Secondly, the very specific security notions
considered for SCC to date do not go far enough to satisfy
the more advanced security requirements in secure network coding, which range from confidentiality to integrity protection of combined data packets in the presence of an active adversary.

 \subsection{Secure index coding}
\label{sub:SIC}

\begin{figure}[htb]
\centering
\begin{tikzpicture}[scale=0.60]
%\draw[step=1cm,gray,very thin] (0,0) grid (11,10);

%the users
\draw (0,2.5) -- (0,3) -- (2,3) -- (2,2.5) -- (0,2.5);
\node at (1, 2.7) {{\scriptsize $U_0$}};
\draw (3,2.5) -- (3,3) -- (5,3) -- (5,2.5) -- (3,2.5);
\node at (4, 2.7) {{\scriptsize $U_1$}};
\draw (8,2.5) -- (8,3) -- (10,3) -- (10,2.5) -- (8,2.5);
\node at (9, 2.7) {{\scriptsize $U_{N-1}$}};

% user-cache
\draw (1,2.5) -- (1,2);
\draw (4,2.5) -- (4,2);
\draw (9,2.5) -- (9,2);

%the caches
\draw (0,1) -- (0,2) -- (2,2) -- (2,1) -- (0,1);
\node at (1,1.5) {{\scriptsize $S_0$}};
\node at (1,0.5) {{\scriptsize $C_0$}};
\draw (3,1) -- (3,2) -- (5,2) -- (5,1) -- (3,1);
\node at (4,1.5) {{\scriptsize $S_1$}};
\node at (4,0.5) {{\scriptsize $C_1$}};
\node at (6.5,2) {$\ldots$};
\draw (8,1) -- (8,2) -- (10,2) -- (10,1) -- (8,1);
\node at (9,1.5) {{\scriptsize $S_{N-1}$}};
\node at (9,0.5) {{\scriptsize $C_{N-1}$}};

%the server
\draw (4,6) -- (4,9) -- (6,9) -- (6,6) --(4,6);
\draw (4,8.5) -- (6,8.5);
\node at (5,8.7) {{\scriptsize $W_0$}};
\draw (4,8) -- (6,8);
\node at (5,8.2) {{\scriptsize $W_1$}};
\draw (4,6.5) -- (6,6.5);
\node at (5,7.5) {$\vdots$};
\node at (5,6.2) {{\scriptsize $W_{N-1}$}};
\node at (3.3, 8) {{\scriptsize Server}};
\node at (3.3, 7.5) {{\scriptsize $S$}};

%broadcast channel
\draw (5,6) -- (5,5.5);
\draw (5,5.5) -- (5,5);
\draw (5,5) -- (1,3);
\draw (5,5) -- (4,3);
\draw (5,5) -- (9,3);

\end{tikzpicture}
  
  \caption{Index coding as coded caching. We denote with $S_i$, where $i \in \{0, .., N-1\}$, the subset of fixed, uncoded messages stored in cache $C_i$.}
  \label{fig:IC}
\end{figure}
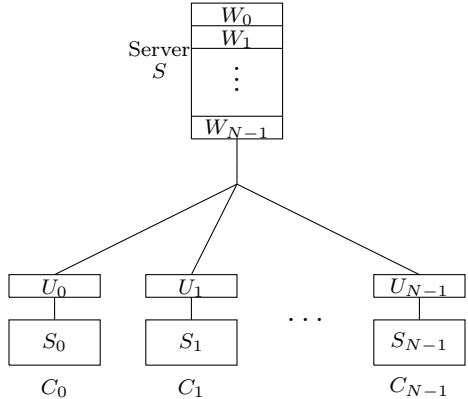

Coded caching is strongly related to index coding \cite{BJK11,BKL10,Lee2015}.\footnote{There is also a strong connection between network coding and index coding, which has already been explored in the literature \cite{Effros2015, Ong2021}.} 
In index coding, a server stores $N$ messages, not necessarily all of the same
size.  Each message is
treated as a single entity and there is no subdivisions.  
The server connects via a broadcast link to users.  Each user has
a local cache which stores some of the messages.   A user requests exactly
one message from the server, and the server broadcasts
some message so that each user's demand is
satisfied.

The relation between coded caching and index coding is well-known and it is explained in \cite{MaddahAli2014,Wan2020}:
the demands in index coding are fixed, and the cache contents are fixed and uncoded, and hence a coded caching
problem is in fact many parallel index coding problems (see Figure \ref{fig:IC}).  Coded caching aims to solve all these problems simultaneously in an efficient manner.

Less explored is the relation between \emph{secure} coded caching and \emph{secure} index coding \cite{Dau2012,Narayanan2018}.
In secure index coding, the adversary is assumed to be able to eavesdrop 
on the server broadcast, and may also have access to some cache content.
The aim is to discover information about the messages.  The literature on this
subject (for example, \cite{Karmoose2017,Mojahedian2017, Ong2016}) generally deals with information-theoretic security,
and techniques proposed include one-time pads and error-correcting codes. From our study, the range of security goals, the adversarial models and the techniques considered in the literature are similar to the SCC literature and have similar gaps.
A conjecture we could make is that secure coded caching can be seen to arise
from gluing together instances of secure index coding, much
as coded caching relates to index coding. If this were to hold,
it could enable advances in secure index coding to translate
into advances in SCC, albeit the issue of efficiency would still hold.

\subsection{Cache-aided single-server private information retrieval}
\label{sub:PIR}

Private information retrieval (PIR) is a cryptographic primitive designed to allow a user to retrieve a file from a database (or server) without revealing which file is being retrieved.
There are many variants of PIR: single server or multiple servers, single user or multiple users, with or without access to side information (caches), and more \cite{Cachin1999,Chor1997,Kazemi2019,Heidarzadeh2019,Heidarzadeh2021,Kushilevitz1997,Li2020}.  Common to all settings is the security concern that a user's request should be hidden from the server(s).  Another variant is that of {\em private information delivery (PID)} \cite{pid}, where, instead, a set of servers decide to supply a file to a user, and wish to hide from the user which server has supplied the file.  In the scenario we consider in this paper we only have a single server, so PID techniques are not directly relevant, but they might be of interest in an extension that considers systems where multiple servers collaborate in supplying content.

In \emph{cache-aided single-server} PIR, the type more relevant to our study, a central server $S$ stores $N$ messages.    
It connects via a
single link to users, each with a local cache.  A cache contains some subset of the $N$ messages, uploaded in a placement phase, which may be coded or uncoded.   Each user requests exactly one message from the server.  A PIR scheme allows the server to broadcast some message to the users in the delivery phase in such a way that the users may retrieve their requested messages without the server learning the identity of these messages.

 In cache-aided PIR, it varies whether information about cache content is or is not a security issue.  For example,  \cite{Tandon2017} has multiple servers and the servers
 know the cache content, and \cite{Kadhe2020} deals with the usual case of private demand as well as the case where both the demand and the cache  content are unknown to the server. 
 
Interestingly, \cite{Vaidya2023} states that the connection between SCC and cache-aided PIR is being investigated, but no reference or detail is provided. It is suggested that functionality-wise the two primitives are similar, but they address two different privacy concerns: PIR aims to protect privacy with respect to the server, while SCC, more specifically SCC schemes that achieve demand privacy, are concerned with users not finding out about other users' demands. 
However, we see no reason why protecting against a curious or even malicious server could not be in the remit of SCC (indeed we consider this a potential model extension in Section \ref{sec:survey}), and we believe a promising line of research could be to more formally explore whether solutions to cache-aided single-server PIR can be directly translated into solutions to achieve coded caching, secure against an untrusted server.

\subsection{Broadcast encryption}
\label{sub:BE}

\begin{figure}[htb]
\centering
\begin{tikzpicture}[scale=0.60]
%\draw[step=1cm,gray,very thin] (0,0) grid (11,10);

%%Server
\draw (4,15.5) -- (4,17) -- (5.5,17) -- (5.5,15.5) -- (4,15.5);
\node at (4.8,16.2) {{\scriptsize $W$}};
\node at (6.7,16.2) {{\scriptsize Server $S$}};
%%links
\draw (4,15.5) -- (0.7,13.5);
\draw (4.5,15.5) -- (2.7,13.5);
\draw (5,15.5) -- (4.7,13.5);
\draw (5.5,15.5) -- (8.7,13.5);
%%Users
\draw (0,13) -- (0,13.5) -- (1.5,13.5) -- (1.5,13) -- (0,13);
\node at (0.8, 13.2) {{\scriptsize $U_0$}};
\draw (2,13) -- (2,13.5) -- (3.5,13.5) -- (3.5,13) -- (2,13);
\node at (2.8, 13.2) {{\scriptsize $U_1$}};
\draw (4,13) -- (4,13.5) -- (5.5,13.5) -- (5.5,13) -- (4,13);
\node at (4.8, 13.2) {{\scriptsize $U_2$}};
\node at (6.8, 13.2){{$\cdots$}};
\draw (8,13) -- (8,13.5) -- (9.5,13.5) -- (9.5,13) -- (8,13);
\node at (8.8, 13.2) {{\scriptsize $U_{K-1}$}};
%%Links
\draw (0.7,12.5) -- (0.7,13);
\draw (2.7,12.5) -- (2.7,13);
\draw (4.7,12.5) -- (4.7,13);
\draw (8.7,12.5) -- (8.7,13);
%%Caches
\draw (0,11.5) -- (0,12.5) -- (1.5,12.5) -- (1.5,11.5) -- (0,11.5);
\node at (0.8, 12) {{\scriptsize $sk_0$}};
\node at (0.8, 11) {{\scriptsize $C_0$}};
\draw (2,11.5) -- (2,12.5) -- (3.5,12.5) -- (3.5,11.5) -- (2,11.5);
\node at (2.8, 12) {{\scriptsize $sk_1$}};
\node at (2.8, 11) {{\scriptsize $C_1$}};
\draw (4,11.5) -- (4,12.5) -- (5.5,12.5) -- (5.5,11.5) -- (4,11.5);
\node at (4.8, 12) {{\scriptsize $sk_2$}};
\node at (4.8, 11) {{\scriptsize $C_2$}};
\node at (6.8, 12){{$\cdots$}};
\draw (8,11.5) -- (8,12.5) -- (9.5,12.5) -- (9.5,11.5) -- (8,11.5);
\node at (8.8, 12) {{\scriptsize $sk_{K-1}$}};
\node at (8.8, 11) {{\scriptsize $C_{K-1}$}};

\node at (5, 10) {{\scriptsize (a) Placement phase}};

%%Server
\draw (4,7) -- (4,8.5) -- (5.5,8.5) -- (5.5,7) -- (4,7);
\node at (4.8,7.7) {{\scriptsize $W$}};
\node at (6.7,7.7) {{\scriptsize Server $S$}};
%%Encryption
\draw (4.8,7)--(4.8,6.5);
\draw (4.8,6) circle (0.5);
\node at (4.8,6) {{\scriptsize $E$}};
%%links
\draw (4.3,6) -- (0.7,3.5);
\draw (4.4,5.7) -- (2.7,3.5);
\draw (4.7,5.5) -- (4.7,3.5);
\draw (5.3,6) -- (8.7,3.5);
%%Users
\draw (0,3) -- (0,3.5) -- (1.5,3.5) -- (1.5,3) -- (0,3);
\node at (0.8, 3.2) {{\scriptsize $U_0$}};
\draw (2,3) -- (2,3.5) -- (3.5,3.5) -- (3.5,3) -- (2,3);
\node at (2.8, 3.2) {{\scriptsize $U_1$}};
\draw (4,3) -- (4,3.5) -- (5.5,3.5) -- (5.5,3) -- (4,3);
\node at (4.8, 3.2) {{\scriptsize $U_2$}};
\node at (6.8, 3.2){{$\cdots$}};
\draw (8,3) -- (8,3.5) -- (9.5,3.5) -- (9.5,3) -- (8,3);
\node at (8.8, 3.2) {{\scriptsize $U_{K-1}$}};
%%links
\draw (0.7,2.5) -- (0.7,3);
\draw (2.7,2.5) -- (2.7,3);
\draw (4.7,2.5) -- (4.7,3);
\draw (8.7,2.5) -- (8.7,3);
%%Caches
\draw (0,1.5) -- (0,2.5) -- (1.5,2.5) -- (1.5,1.5) -- (0,1.5);
\node at (0.8, 2) {{\scriptsize $\times$}};
%\node at (0.8, 2) {{\scriptsize {\ding{55}}}};
\node at (0.8, 1) {{\scriptsize $C_0$}};
\draw (2,1.5) -- (2,2.5) -- (3.5,2.5) -- (3.5,1.5) -- (2,1.5);
\node at (2.8, 2) {{\scriptsize $\checkmark$}};
%\node at (2.8, 2) {{\scriptsize {\ding{51}}}};
\node at (2.8, 1) {{\scriptsize $C_1$}};
\draw (4,1.5) -- (4,2.5) -- (5.5,2.5) -- (5.5,1.5) -- (4,1.5);
\node at (4.8, 2) {{\scriptsize $\checkmark$}};
%\node at (4.8, 2) {{\scriptsize {\ding{51}}}};
\node at (4.8, 1) {{\scriptsize $C_2$}};
\node at (6.8, 2){{$\cdots$}};
\draw (8,1.5) -- (8,2.5) -- (9.5,2.5) -- (9.5,1.5) -- (8,1.5);
\node at (8.8, 2) {{\scriptsize $\times$}};
%\node at (8.8, 2) {{\scriptsize {\ding{55}}}};
\node at (8.8, 1) {{\scriptsize $C_{K-1}$}};

\node at (5, 0) {{\scriptsize (b) Delivery phase}};

\end{tikzpicture}
  
  \caption{BE as coded caching. 
  In this example, file $W$ is encrypted for target set $\{1, 2\}$. A $\checkmark$ indicates that the decryption using the stored secret key yields the correct file.}
  \label{fig:BE}
\end{figure}

Broadcast Encryption (BE) is a primitive designed to efficiently deliver encrypted content to a set of target users. BE is a well-established research area in cryptography. 
While
earlier schemes used combinatorial techniques in a symmetric
cryptography setting \cite{FiatNaor1994}, these ideas have since been considered
formally in a public-key context, and precise security
models have been established (e.g., \cite{GentryW09,Agrawal020}). This has led to the
development of extensions that satisfy further sophisticated
security requirements, such as ensuring the anonymity of users
within the target set \cite{LibertPQ12}.

Here, we reflect on the relation between BE and SCC as these primitives are both concerned with the efficient delivery of content.
On the one hand, we could view BE as a very basic realisation of coded caching (see Figure \ref{fig:BE}). Indeed, in the placement phase secret keys would be placed in the caches (Figure \ref{fig:BE}(a)), i.e., the caches would not contain any content other than key material. In the delivery phase (Figure \ref{fig:BE}(b)), the same encrypted content would be broadcast to all caches, but only the caches with an appropriate secret key, namely a key corresponding to a user in the broadcast target set, are able to recover the content by decryption.\footnote{We note that in this representation we are omitting the demand phase, which is not currently modelled in either BE or SCC.} Intuitively, the security properties of the BE scheme would translate to content confidentiality and file privacy for the resulting SCC scheme. Speculatively, anonymous BE could potentially offer protection with respect to demand privacy.
The projected security benefits of this approach however are undermined by the complete absence of efficiency gain derived from pre-populating the caches, defeating the overall goal of  reducing the load on the network during peak times of content delivery.

On the other hand, let's consider SCC as a building block to achieve BE. Functionality-wise, it seems clear that coded caching can implement what BE requires, i.e., the efficient delivery of content to a set of users (caches). It actually provides a more fine-grained solution in that with the same broadcast, users can access different content according to how their caches were pre-populated.
The challenge is that from a security standpoint current SCC solutions do not deliver the level of security required by BE, and the gap to bridge is, from the literature we have reviewed, very large.

Our analysis of SCC in the context of related primitives, {visualised in Table \ref{tab:primitives},} highlights several overlaps and possible interesting research questions, which we summarize in Section \ref{Open}.
Importantly, we note that the lack of formalisation of an SCC scheme means that proofs of formal relations and security reductions cannot currently be provided. Indeed, highlighting this is yet another incentive to address the formalisation shortcoming in the area.

\begin{table}[h!]
	\centering
	\scriptsize % smaller font
	\renewcommand{\arraystretch}{1.3} % more row spacing
	\begin{tabular}{|p{2cm}|p{3cm}|p{3cm}|p{2.5cm}|p{3.2cm}|}
		\hline\hline
		\textbf{Primitive} & \textbf{Functionality and goal} & \textbf{Security goals \newline and adversaries} & \textbf{Relation to SCC \newline (functionality-wise)} & \textbf{Fundamental challenges \& directions for future work} \\
		\hline\hline
		
		Secure Network Coding (SNC) & 
		1 source emitting $N$ messages to multiple receivers wanting a subset of messages via intermediate nodes &
		Message confidentiality (eavesdroppers and Byzantine adversaries) &
		SNC $\Rightarrow$ SCC &
		- User demand model not considered
		
		- Only one instance of placement-delivery cycle is considered in SNC \newline
		\\
		\hline
		
		Secure Index Coding (SIC)&
		1 source storing $N$ messages, local caches storing some of the messages, receivers requesting 1 message (single instance of CC) &
		Message confidentiality (eavesdropper with potential access to cache content) &
		SCC = many instances of SIC &
		- Combination of multiple SICs is currently inefficient \newline

		\faLightbulbO \ \ Find a way to combine multiple SICs efficiently \\
		\hline
		
		Cache-aided PIR (CA-PIR) &
		Same as SCC &
		Demand privacy (curious/malicious server) &
		CA-PIR $\iff$ SCC \newline
		&
		\faLightbulbO \ \ Consider privacy wrt a curious/malicious server in SCC \newline
		\faLightbulbO \ \ Formally prove equi\-valence \\
		\hline
		
		Broadcast Encryption (BE) &
		1 source efficiently broadcasting 1 message to a subset $S$ of $N$ users &
		Message confidentiality,
		
		Anonymity of users in $S$ \newline
		(active adversaries) &
		BE $\Rightarrow$ SCC (keys in cache) \newline
		SCC $\Rightarrow$ BE &
		- Inefficient construction from BE to SCC \newline
		- Inadequate security modelling for SCC \newline
		
		\faLightbulbO \ \ Provide better security modelling for SCC to close the gap \\
		\hline\hline
	\end{tabular}
	\caption{{Overview of secure content delivery primitives, their goals, relation to Secure Coded Caching (SCC), challenges (denoted with -), and future directions of work (denoted with \faLightbulbO)}.}
	\label{tab:primitives}
\end{table}

\section{Open problems and research directions}\label{Open}
As a result of our systematisation of knowledge in the area of secure coded caching, we identify the following open problems and research directions.

\subsection{A stronger and more realistic security model}
\label{sec:strongermodel}

As discussed in earlier sections, there seems to be a lack of a clearly established and fully developed model for secure coded caching. 
Most proposals focus only on the security of individual parts of the process, and different proposals consider different network models under different assumptions, making it difficult to compare or combine solutions in secure ways. A comprehensive model can be crucial in capturing more advanced attack scenarios, and would allow a systematic design and evaluation of security solutions at desired levels of real-world security.  In particular, we focus on two themes discussed in Section \ref{subsec:analysisanddiscussion} that should be addressed in an integrated manner:

\begin{description}

\item[Capabilities of adversaries]

So far the main adversaries considered have been \emph{passive} attackers, whose capabilities are restricted to eavesdropping on communications and observing transcripts
without deviating from prescribed behaviour. While analytically convenient, this
abstraction is insufficient for realistic deployment settings.  In practice, for real-world attack scenarios, one should consider security against \emph{active} attackers as well, who may have the ability to arbitrarily deviate from the protocol specification. Such adversaries may 
be able to modify communication, corrupt information in storage, and corrupt users (which may include the server) of the system.

\item[Security goals]

 In tandem with extending the security model to include more capable adversaries, the security objectives must also be correspondingly strengthened.
In addition to content confidentiality and demand privacy, the model should explicitly incorporate formal notions of integrity and authentication, applying both
to user demands and to the broadcast transmissions generated during the delivery
phase.
\end{description}

We note that if SCC is to be deployable in the real-world, these issues cannot be addressed in a piecewise fashion.  An integrated approach will allow formalisation of security properties and formal security proofs.  

\subsection{Diversification and modernisation of techniques}
\label{sec:moderntechniques}

In systematising the current literature in secure coded caching, it became apparent that the overwhelmingly used technique to achieve security is the one-time pad.  While one-time pads achieve perfect secrecy in a restricted model, this method does come with many practical limitations, as discussed in Section \ref{subsec:analysisanddiscussion}.  
We summarise two research directions arising from the discussion:

\begin{description}

\item[Key management]

It is well-known that the one-time pad has many practical issues associated with it. 
The restrictive adversarial model and security goal are discussed in Section \ref{sec:survey}.  One of the central issues is that of key management, which becomes particularly acute in the coded caching setting due to the multiplicity of users, the presence of cache placement and delivery
phases, and the potential for dynamic participation. Information-theoretic masking
requires secret keys whose entropy matches that of the protected content, leading to
linear key-length overhead and necessitating secure generation, distribution, storage, and periodic refresh. In large-scale or heterogeneous networks, this introduces
significant scalability constraints.  Indeed, even if we should use other cryptographic techniques, key lifecycle management is still a fundamental consideration, and its practicality will have direct impact on whether a system can be sustainable, and this has not been discussed in any detail in the literature.   Techniques of key distribution or key pre-distribution may be usefully investigated here.

\item[Computational security]

The one-time pad provides \emph{unconditional security},  where
no limit is placed on the computational power of an attacker.  However, in a practical scenario, the attacker is restricted to computations that
can be carried out within a reasonable amount of time. This setting is largely unexplored in the SCC literature. This is an area where modern cryptographic techniques can deliver feasible
solutions unattainable by information-theoretic approaches alone. Formal cryptographic definitions and security models to capture what an SCC scheme requires in practice would be an interesting and useful direction of research.

An example of such an approach in a related setting is provided by Vo, Lai, Yuan, Nepal and Li's very recent work on privacy in content delivery networks \cite{oblivCDN}.  Here they propose the use of Trusted Execution Environments and range oblivious RAM \cite{chakraborti2019rOram} to achieve what is referred to as video confidentiality, akin to file privacy, and access pattern protection, which aims at hiding the requests from the server (analogous to demand privacy). Their work features a more sophisticated threat model than those appearing in the coded caching literature and addresses privacy considerations using more modern techniques that are practical for the specific application setting.  Three (complementary) research directions we suggest in relation to this work are to explore the extent to which coded caching gains can be exploited directly in such a setting, to investigate whether these privacy models translate to a coded caching context or whether they require modification and, similarly, to examine the potential of tools such as range oblivious RAM to provide privacy in a coded caching system.   
\end{description}

\subsection{Secure content delivery primitives ecosystem} %Content delivery primitives ecosystem
\label{sec:ecosystem}

 As we discussed in Section \ref{sec:related}, SCC has intersections with many other primitives designed for the delivery of content.  We summarise some of the open questions that arose in the discussion:
 
 \begin{enumerate}
   \item \textbf{SCC and secure index coding}
  Can secure coded caching be achieved by combining instances of secure index coding? 
 \item \textbf{SCC and cache-aided single-server PIR}
If SCC is extended to provide privacy with respect to the server, what is the exact relationship between cache-aided single-server PIR and SCC? 
 \item \textbf{SCC and broadcast encryption}
  While SCC does not deliver the level of security required by broadcast encryption, as it currently stands, it does provide an efficient delivery of content.  Can better security provision in SCC close this gap?
  \end{enumerate}
  
The overarching question is, by extending the adversarial model and security goals of SCC, will we be able to have an accurate picture of the secure content delivery ecosystem, showing how SCC relates to other primitives in this area?

\section{Conclusions}\label{sec:conc}

We have highlighted a gap between the cutting-edge coding techniques proposed in the area of secure coded caching to help maximise performance (indeed, this is the key to the success of coded caching and arguably its driving force), and the cryptographic techniques and practical considerations adopted to address security concerns.  {Our survey establishes the groundwork of contextualising the problem and the limitations, showing that appropriate modelling is not straightforward and much research is needed in this direction.   In particular, one of the main findings in our survey is the limited techniques proposed in the literature, and we have proposed an avenue of study into other security primitives that can potentially be integrated with coded caching.}   We believe bridging this gap and modernising the approach to security would be highly beneficial to this area, especially given its ambition to address real-world requirements from modern networks.

\bibliographystyle{elsarticle-num}
%\bibliography{securecaching}

\begin{thebibliography}{10}
\expandafter\ifx\csname url\endcsname\relax
  \def\url#1{\texttt{#1}}\fi
\expandafter\ifx\csname urlprefix\endcsname\relax\def\urlprefix{URL }\fi
\expandafter\ifx\csname href\endcsname\relax
  \def\href#1#2{#2} \def\path#1{#1}\fi

\bibitem{MaddahAli2014}
M.~A. {Maddah-Ali}, U.~Niesen, Fundamental limits of caching, IEEE Transactions
  on Information Theory 60~(5) (2014) 2856--2867.
\newblock \href {https://doi.org/10.1109/TIT.2014.2306938}
  {\path{doi:10.1109/TIT.2014.2306938}}.

\bibitem{Yan2017}
Q.~Yan, M.~Cheng, X.~Tang, Q.~Chen, On the placement delivery array design for
  centralized coded caching scheme, IEEE Transactions on Information Theory
  63~(9) (2017) 5821--5833.
\newblock \href {https://doi.org/10.1109/TIT.2017.2725272}
  {\path{doi:10.1109/TIT.2017.2725272}}.

\bibitem{Chittoor2021}
H.~H.~S. Chittoor, P.~Krishnan, K.~V.~S. Sree, B.~Mamillapalli, Subexponential
  and linear subpacketization coded caching via projective geometry, IEEE
  Transactions on Information Theory 67~(9) (2021) 6193--6222.
\newblock \href {https://doi.org/10.1109/TIT.2021.3095471}
  {\path{doi:10.1109/TIT.2021.3095471}}.

\bibitem{Cheng2021}
M.~Cheng, J.~Wang, X.~Zhong, Q.~Wang, A framework of constructing placement
  delivery arrays for centralized coded caching, IEEE Transactions on
  Information Theory 67~(11) (2021) 7121--7131.
\newblock \href {https://doi.org/10.1109/TIT.2021.3112492}
  {\path{doi:10.1109/TIT.2021.3112492}}.

\bibitem{MaddahAli2015}
M.~A. Maddah-Ali, U.~Niesen, Decentralized coded caching attains order-optimal
  memory-rate tradeoff, IEEE/ACM Transactions on Networking 23~(4) (2015)
  1029--1040.
\newblock \href {https://doi.org/10.1109/TNET.2014.2317316}
  {\path{doi:10.1109/TNET.2014.2317316}}.

\bibitem{Shariatpanahi2016}
S.~P. Shariatpanahi, S.~A. Motahari, B.~H. Khalaj, Multi-server coded caching,
  IEEE Transactions on Information Theory 62~(12) (2016) 7253--7271.
\newblock \href {https://doi.org/10.1109/TIT.2016.2614722}
  {\path{doi:10.1109/TIT.2016.2614722}}.

\bibitem{Karamchandani2016}
N.~Karamchandani, U.~Niesen, M.~A. Maddah-Ali, S.~N. Diggavi, Hierarchical
  coded caching, IEEE Transactions on Information Theory 62~(6) (2016)
  3212--3229.
\newblock \href {https://doi.org/10.1109/TIT.2016.2557804}
  {\path{doi:10.1109/TIT.2016.2557804}}.

\bibitem{YanParampalli2017}
Q.~Yan, U.~Parampalli, X.~Tang, Q.~Chen, Online coded caching with random
  access, IEEE Communications Letters 21~(3) (2017) 552--555.
\newblock \href {https://doi.org/10.1109/LCOMM.2016.2631552}
  {\path{doi:10.1109/LCOMM.2016.2631552}}.

\bibitem{Lampiris2018}
E.~Lampiris, P.~Elia, Adding transmitters dramatically boosts coded-caching
  gains for finite file sizes, IEEE Journal on Selected Areas in Communications
  36~(6) (2018) 1176--1188.
\newblock \href {https://doi.org/10.1109/JSAC.2018.2844960}
  {\path{doi:10.1109/JSAC.2018.2844960}}.

\bibitem{KrishnanNamboodiri2022}
K.~K. {Krishnan Namboodiri}, B.~{Sundar Rajan}, Multi-access coded caching with
  coded placement, in: 2022 IEEE Wireless Communications and Networking
  Conference (WCNC), 2022, pp. 2274--2279.
\newblock \href {https://doi.org/10.1109/WCNC51071.2022.9771891}
  {\path{doi:10.1109/WCNC51071.2022.9771891}}.

\bibitem{KrishnanNamboodiriPeter2022}
K.~K. {Krishnan Namboodiri}, E.~Peter, B.~{Sundar Rajan}, Extended placement
  delivery arrays for multi-antenna coded caching scheme, in: 2022 IEEE
  International Symposium on Information Theory (ISIT), 2022, pp. 1518--1523.
\newblock \href {https://doi.org/10.1109/ISIT50566.2022.9834422}
  {\path{doi:10.1109/ISIT50566.2022.9834422}}.

\bibitem{Das2022}
N.~Das, B.~{Sundar Rajan}, A shared cache coded caching scheme using designs
  and circuits of matrices, in: 2023 IEEE Information Theory Workshop (ITW),
  2023, pp. 131--135.

\bibitem{fastly}
A.~Hern, Massive internet outage hits websites including {A}mazon, gov.uk and
  {G}uardian, {https://tinyurl.com/fastlyoutage}, accessed 6 August 2021
  (2021).

\bibitem{oblivCDN}
V.~Vo, S.~Lai, X.~Yuan, S.~Nepal, Q.~Li, Obliv{CDN}: A practical
  privacy-preserving {CDN} with oblivious content access, in: 20th ACM Asia
  Conference on Computer and Communications Security (ASIACCS), Ha Noi,
  Vietnam, 2025.
\newblock \href {https://doi.org/10.48550/arXiv.2501.07262}
  {\path{doi:10.48550/arXiv.2501.07262}}.

\bibitem{Sengupta2015}
A.~Sengupta, R.~Tandon, T.~C. Clancy, Fundamental limits of caching with secure
  delivery, IEEE Transactions on Information Forensics and Security 10~(2)
  (2015) 355--370.
\newblock \href {https://doi.org/10.1109/TIFS.2014.2375553}
  {\path{doi:10.1109/TIFS.2014.2375553}}.

\bibitem{Abdullahi2015}
I.~Abdullahi, S.~Arif, S.~Hassan, Survey on caching approaches in information
  centric networking, Journal of Network and Computer Applications 56 (2015)
  48–59.
\newblock \href {https://doi.org/10.1016/j.jnca.2015.06.011}
  {\path{doi:10.1016/j.jnca.2015.06.011}}.

\bibitem{Li2018}
L.~Li, G.~Zhao, R.~S. Blum, A survey of caching techniques in cellular
  networks: Research issues and challenges in content placement and delivery
  strategies, IEEE Communications Surveys \& Tutorials 20~(3) (2018)
  1710--1732.
\newblock \href {https://doi.org/10.1109/COMST.2018.2820021}
  {\path{doi:10.1109/COMST.2018.2820021}}.

\bibitem{Yao2019}
J.~Yao, T.~Han, N.~Ansari, On mobile edge caching, IEEE Communications Surveys
  \& Tutorials 21~(3) (2019) 2525--2553.
\newblock \href {https://doi.org/10.1109/COMST.2019.2908280}
  {\path{doi:10.1109/COMST.2019.2908280}}.

\bibitem{Ghaznavi2021}
M.~Ghaznavi, E.~Jalalpour, M.~A. Salahuddin, R.~Boutaba, D.~Migault, S.~Preda,
  Content delivery network security: A survey, IEEE Communications Surveys \&
  Tutorials 23~(4) (2021) 2166--2190.
\newblock \href {https://doi.org/10.1109/COMST.2021.3093492}
  {\path{doi:10.1109/COMST.2021.3093492}}.

\bibitem{Pruthvi2023}
C.~N. Pruthvi, H.~S. Vimala, S.~J., A systematic survey on content caching in
  {ICN} and {ICN-IoT}: Challenges, approaches and strategies, Computer Networks
  233 (2023) 109896.
\newblock \href {https://doi.org/10.1016/j.comnet.2023.109896}
  {\path{doi:10.1016/j.comnet.2023.109896}}.

\bibitem{Khan2024}
Y.~Khan, S.~Mustafa, R.~W. Ahmad, T.~Maqsood, F.~Rehman, J.~Ali, J.~J.
  Rodrigues, Content caching in mobile edge computing: a survey, Cluster
  Computing 27 (2024) 8817–8864.
\newblock \href {https://doi.org/10.1007/s10586-024-04459-7}
  {\path{doi:10.1007/s10586-024-04459-7}}.

\bibitem{Li2024}
H.~Li, M.~Sun, F.~Xia, X.~Xu, M.~Bilal, A survey of edge caching: Key issues
  and challenges, Tsinghua Science and Technology 29~(3) (2024) 818--842.
\newblock \href {https://doi.org/10.26599/TST.2023.9010051}
  {\path{doi:10.26599/TST.2023.9010051}}.

\bibitem{Nguyen2023}
T.-V. Nguyen, A.-T. Tran, N.-N. Dao, H.~Moon, S.~Cho, Information fusion on
  delivery: A survey on the roles of mobile edge caching systems, Information
  Fusion 89 (2023) 486--509.
\newblock \href {https://doi.org/10.1016/j.inffus.2022.08.029}
  {\path{doi:10.1016/j.inffus.2022.08.029}}.

\bibitem{Zhang2025}
H.~Zhang, J.~Wang, Z.~Zhao, Z.~Zhao, A survey of edge caching security:
  Framework, methods, and challenges, Journal of Systems Architecture 158
  (2025) 103306.
\newblock \href {https://doi.org/10.1016/j.sysarc.2024.103306}
  {\path{doi:10.1016/j.sysarc.2024.103306}}.

\bibitem{XZhang2025}
X.~Zhang, Y.~Zhou, D.~Wu, Q.~Z. Sheng, S.~Riaz, M.~Hu, L.~Xiao, A survey on
  privacy-preserving caching at network edge: Classification, solutions, and
  challenges, ACM Comput. Surv. 57~(5) (Jan. 2025).
\newblock \href {https://doi.org/10.1145/3706630} {\path{doi:10.1145/3706630}}.

\bibitem{Barrios2023}
C.~Barrios, M.~Kumar, Service caching and computation reuse strategies at the
  edge: A survey, ACM Comput. Surv. 56~(2) (Sep. 2023).
\newblock \href {https://doi.org/10.1145/3609504} {\path{doi:10.1145/3609504}}.

\bibitem{Ng2021}
J.~S. Ng, W.~Y.~B. Lim, N.~C. Luong, Z.~Xiong, A.~Asheralieva, D.~Niyato,
  C.~Leung, C.~Miao, A comprehensive survey on coded distributed computing:
  Fundamentals, challenges, and networking applications, IEEE Communications
  Surveys \& Tutorials 23~(3) (2021) 1800--1837.
\newblock \href {https://doi.org/10.1109/COMST.2021.3091684}
  {\path{doi:10.1109/COMST.2021.3091684}}.

\bibitem{Naderializadeh2017}
N.~Naderializadeh, M.~A. Maddah-Ali, A.~S. Avestimehr, On the optimality of
  separation between caching and delivery in general cache networks, in: 2017
  IEEE International Symposium on Information Theory (ISIT), 2017, pp.
  1232--1236.
\newblock \href {https://doi.org/10.1109/ISIT.2017.8006725}
  {\path{doi:10.1109/ISIT.2017.8006725}}.

\bibitem{Ravindrakumar2016}
V.~Ravindrakumar, P.~Panda, N.~Karamchandani, V.~Prabhakaran, Fundamental
  limits of secretive coded caching, in: 2016 IEEE International Symposium on
  Information Theory (ISIT), 2016, pp. 425--429.
\newblock \href {https://doi.org/10.1109/ISIT.2016.7541334}
  {\path{doi:10.1109/ISIT.2016.7541334}}.

\bibitem{Gurjarpadhye2021}
C.~Gurjarpadhye, J.~Ravi, S.~Kamath, B.~K. Dey, N.~Karamchandani, Fundamental
  limits of demand-private coded caching, IEEE Transactions on Information
  Theory 68~(6) (2022) 4106--4134.
\newblock \href {https://doi.org/10.1109/TIT.2022.3150336}
  {\path{doi:10.1109/TIT.2022.3150336}}.

\bibitem{Pedarsani2016}
R.~Pedarsani, M.~A. Maddah-Ali, U.~Niesen, Online coded caching, IEEE/ACM
  Trans. Netw. 24~(2) (2016) 836–845.

\bibitem{Birk1998}
Y.~Birk, T.~Kol, Informed-source coding-on-demand ({ISCOD}) over broadcast
  channels, in: Proceedings. IEEE INFOCOM '98, the Conference on Computer
  Communications. Seventeenth Annual Joint Conference of the IEEE Computer and
  Communications Societies. Gateway to the 21st Century (Cat. No.98, Vol.~3,
  1998, pp. 1257--1264.
\newblock \href {https://doi.org/10.1109/INFCOM.1998.662940}
  {\path{doi:10.1109/INFCOM.1998.662940}}.

\bibitem{Shariatpanahi2017}
S.~P. Shariatpanahi, G.~Caire, B.~H. Khalaj, Multi-antenna coded caching, in:
  2017 IEEE International Symposium on Information Theory (ISIT), 2017, pp.
  2113--2117.
\newblock \href {https://doi.org/10.1109/ISIT.2017.8006902}
  {\path{doi:10.1109/ISIT.2017.8006902}}.

\bibitem{Zewail2016}
A.~A. Zewail, A.~Yener, Fundamental limits of secure device-to-device coded
  caching, in: 2016 50th Asilomar Conference on Signals, Systems and Computers,
  2016, pp. 1414--1418.
\newblock \href {https://doi.org/10.1109/ACSSC.2016.7869609}
  {\path{doi:10.1109/ACSSC.2016.7869609}}.

\bibitem{Wan2022}
K.~Wan, M.~Cheng, D.~Liang, G.~Caire, Multiaccess coded caching with private
  demands, in: 2022 IEEE International Symposium on Information Theory (ISIT),
  2022, pp. 1390--1395.
\newblock \href {https://doi.org/10.1109/ISIT50566.2022.9834460}
  {\path{doi:10.1109/ISIT50566.2022.9834460}}.

\bibitem{Peter2022}
E.~Peter, K.~K. Krishnan~Namboodiri, B.~{Sundar Rajan}, Shared cache coded
  caching schemes with known user-to-cache association profile using placement
  delivery arrays, in: 2022 IEEE Information Theory Workshop (ITW), 2022, pp.
  678--683.
\newblock \href {https://doi.org/10.1109/ITW54588.2022.9965896}
  {\path{doi:10.1109/ITW54588.2022.9965896}}.

\bibitem{Welsh1988}
D.~J.~A. Welsh, Codes and cryptography, Oxford University Press, Oxford, 1988.

\bibitem{Qi2022}
C.~Qi, J.~Ravi, Coded caching with file and demand privacy, IEEE Communications
  Letters 26~(9) (2022) 1979--1983.
\newblock \href {https://doi.org/10.1109/LCOMM.2022.3184765}
  {\path{doi:10.1109/LCOMM.2022.3184765}}.

\bibitem{Ravindrakumar2018}
V.~Ravindrakumar, P.~Panda, N.~Karamchandani, V.~M. Prabhakaran, Private coded
  caching, IEEE Transactions on Information Forensics and Security 13~(3)
  (2018) 685--694.
\newblock \href {https://doi.org/10.1109/TIFS.2017.2765503}
  {\path{doi:10.1109/TIFS.2017.2765503}}.

\bibitem{ShresthaPIMRC2021}
S.~S. Meel, B.~{Sundar Rajan}, Secretive coded caching from {PDA}s, in: 2021
  IEEE 32nd Annual International Symposium on Personal, Indoor and Mobile Radio
  Communications (PIMRC), 2021, pp. 373--379.
\newblock \href {https://doi.org/10.1109/PIMRC50174.2021.9569455}
  {\path{doi:10.1109/PIMRC50174.2021.9569455}}.

\bibitem{Shrestha2021}
S.~S. Meel, B.~{Sundar Rajan}, Secretive coded caching with shared caches, IEEE
  Communications Letters 25~(9) (2021) 2849--2853.
\newblock \href {https://doi.org/10.1109/LCOMM.2021.3094405}
  {\path{doi:10.1109/LCOMM.2021.3094405}}.

\bibitem{PeterISIT2022}
E.~Peter, K.~K. {Krishnan Namboodiri}, B.~{Sundar Rajan}, A secretive coded
  caching for shared cache systems using placement delivery arrays, in: 2022
  IEEE International Symposium on Information Theory (ISIT), 2022, pp.
  1402--1407.
\newblock \href {https://doi.org/10.1109/ISIT50566.2022.9834824}
  {\path{doi:10.1109/ISIT50566.2022.9834824}}.

\bibitem{Suthan2017}
C.~H.~H. Suthan, I.~Chugh, P.~Krishnan, An improved secretive coded caching
  scheme exploiting common demands, in: 2017 IEEE Information Theory Workshop
  (ITW), 2017, pp. 66--70.
\newblock \href {https://doi.org/10.1109/ITW.2017.8277998}
  {\path{doi:10.1109/ITW.2017.8277998}}.

\bibitem{Wan2021}
K.~Wan, G.~Caire, On coded caching with private demands, IEEE Transactions on
  Information Theory 67~(1) (2021) 358--372.
\newblock \href {https://doi.org/10.1109/TIT.2020.3036313}
  {\path{doi:10.1109/TIT.2020.3036313}}.

\bibitem{kamath2019demand}
S.~Kamath, Demand private coded caching, CoRR abs/1909.03324 (2019).
\newblock \href {https://doi.org/10.48550/arXiv.1909.03324}
  {\path{doi:10.48550/arXiv.1909.03324}}.

\bibitem{Kamath2020}
S.~Kamath, J.~Ravi, B.~K. Dey, Demand-private coded caching and the exact
  trade-off for $n=k=2$, in: 2020 National Conference on Communications (NCC),
  2020, pp. 1--6.
\newblock \href {https://doi.org/10.1109/NCC48643.2020.9056026}
  {\path{doi:10.1109/NCC48643.2020.9056026}}.

\bibitem{Gurjarpadhye20}
C.~Gurjarpadhye, J.~Ravi, B.~K. Dey, N.~Karamchandani, Improved memory-rate
  trade-off for caching with demand privacy, in: 2020 IEEE Information Theory
  Workshop (ITW), 2021, pp. 1--5.
\newblock \href {https://doi.org/10.1109/ITW46852.2021.9457647}
  {\path{doi:10.1109/ITW46852.2021.9457647}}.

\bibitem{Namboodiri2022privacy}
K.~K. Krishnan~Namboodiri, B.~Sundar~Rajan, Multi-access coded caching with
  demand privacy, in: 2022 IEEE Wireless Communications and Networking
  Conference (WCNC), 2022, pp. 2280--2285.
\newblock \href {https://doi.org/10.1109/WCNC51071.2022.9771663}
  {\path{doi:10.1109/WCNC51071.2022.9771663}}.

\bibitem{Aravind2020}
V.~R. Aravind, P.~K. Sarvepalli, A.~Thangaraj, Subpacketization in coded
  caching with demand privacy, in: 2020 National Conference on Communications
  (NCC), 2020, pp. 1--6.
\newblock \href {https://doi.org/10.1109/NCC48643.2020.9055996}
  {\path{doi:10.1109/NCC48643.2020.9055996}}.

\bibitem{Cheng2020}
M.~Cheng, D.~Liang, R.~Wei, On secure coded caching via combinatorial method,
  CoRR abs/2005.01043 (2020).
\newblock \href {https://doi.org/10.48550/arXiv.2005.01043}
  {\path{doi:10.48550/arXiv.2005.01043}}.

\bibitem{Gholami2022}
A.~Gholami, K.~Wan, H.~Sun, M.~Ji, G.~Caire, Coded caching with private demands
  and caches, in: 2022 IEEE International Symposium on Information Theory
  (ISIT), 2022, pp. 1396--1401.
\newblock \href {https://doi.org/10.1109/ISIT50566.2022.9834846}
  {\path{doi:10.1109/ISIT50566.2022.9834846}}.

\bibitem{Ma2021}
K.~Ma, S.~Shao, J.~Shao, Secure coded caching with colluding users, in: 2021
  Computing, Communications and IoT Applications (ComComAp), 2021, pp.
  329--334.
\newblock \href {https://doi.org/10.1109/ComComAp53641.2021.9653057}
  {\path{doi:10.1109/ComComAp53641.2021.9653057}}.

\bibitem{Yan2021}
Q.~Yan, D.~Tuninetti, Fundamental limits of caching for demand privacy against
  colluding users, IEEE Journal on Selected Areas in Information Theory 2~(1)
  (2021) 192--207.
\newblock \href {https://doi.org/10.1109/JSAIT.2021.3053372}
  {\path{doi:10.1109/JSAIT.2021.3053372}}.

\bibitem{Namboodiri2021}
K.~{Krishnan Namboodiri}, B.~Sundar~Rajan, Optimal demand private coded caching
  for users with small buffers, in: 2021 IEEE International Symposium on
  Information Theory (ISIT), 2021, pp. 706--711.
\newblock \href {https://doi.org/10.1109/ISIT45174.2021.9518070}
  {\path{doi:10.1109/ISIT45174.2021.9518070}}.

\bibitem{mceliecesarwate}
R.~J. McEliece, D.~V. Sarwate, \href{https://doi.org/10.1145/358746.358762}{On
  sharing secrets and reed-solomon codes}, Commun. ACM 24~(9) (1981) 583–584.
\newblock \href {https://doi.org/10.1145/358746.358762}
  {\path{doi:10.1145/358746.358762}}.
\newline\urlprefix\url{https://doi.org/10.1145/358746.358762}

\bibitem{Engelmann2017}
F.~Engelmann, E.~Elia, A content-delivery protocol, exploiting the privacy
  benefits of coded caching, in: 2017 15th International Symposium on Modeling
  and Optimization in Mobile, Ad Hoc, and Wireless Networks (WiOpt), 2017, pp.
  1--6.
\newblock \href {https://doi.org/10.23919/WIOPT.2017.7959863}
  {\path{doi:10.23919/WIOPT.2017.7959863}}.

\bibitem{LPQS17}
J.~Leguay, G.~S. Paschos, E.~A. Quaglia, B.~Smyth, Crypto{C}ache: Network
  caching with confidentiality, in: {IEEE} International Conference on
  Communications ({ICC}), {IEEE}, Paris, 2017, pp. 1--6.
\newblock \href {https://doi.org/10.1109/ICC.2017.7996866}
  {\path{doi:10.1109/ICC.2017.7996866}}.

\bibitem{Ahlswede2000}
R.~Ahlswede, N.~Cai, S.-Y.~R. Li, R.~W. Yeung, Network information flow, IEEE
  Transactions on Information Theory 46~(4) (2000) 1204--1216.
\newblock \href {https://doi.org/10.1109/18.850663}
  {\path{doi:10.1109/18.850663}}.

\bibitem{Cai2011}
N.~Cai, T.~Chan, Theory of secure network coding, Proceedings of the IEEE
  99~(3) (2011) 421--437.
\newblock \href {https://doi.org/10.1109/JPROC.2010.2094592}
  {\path{doi:10.1109/JPROC.2010.2094592}}.

\bibitem{Liu2015}
G.~Liu, B.~Liu, X.~Liu, F.~Li, W.~Guo, Low-complexity secure network coding
  against wiretapping using intra/inter-generation coding, China Communications
  12~(6) (2015) 116--125.
\newblock \href {https://doi.org/10.1109/CC.2015.7122470}
  {\path{doi:10.1109/CC.2015.7122470}}.

\bibitem{Wang2019}
X.~Wang, N.~Qin, Y.~Liu, A secure network coding system against wiretap
  attacks, in: 2019 34rd Youth Academic Annual Conference of Chinese
  Association of Automation (YAC), 2019, pp. 62--67.
\newblock \href {https://doi.org/10.1109/YAC.2019.8787724}
  {\path{doi:10.1109/YAC.2019.8787724}}.

\bibitem{Ho2004}
T.~Ho, B.~Leong, R.~Koetter, M.~Medard, M.~Effros, D.~R. Karger, Byzantine
  modification detection in multicast networks using randomized network coding,
  in: International Symposium on Information Theory, 2004. ISIT 2004.
  Proceedings., 2004, p. 143.
\newblock \href {https://doi.org/10.1109/ISIT.2004.1365180}
  {\path{doi:10.1109/ISIT.2004.1365180}}.

\bibitem{Charles2006}
D.~Charles, K.~Jain, K.~Lauter, Signatures for network coding, in: 2006 40th
  Annual Conference on Information Sciences and Systems, 2006, pp. 857--863.
\newblock \href {https://doi.org/10.1109/CISS.2006.286587}
  {\path{doi:10.1109/CISS.2006.286587}}.

\bibitem{Zhao2007}
F.~Zhao, T.~Kalker, M.~Medard, K.~J. Han, Signatures for content distribution
  with network coding, in: 2007 IEEE International Symposium on Information
  Theory, 2007, pp. 556--560.
\newblock \href {https://doi.org/10.1109/ISIT.2007.4557283}
  {\path{doi:10.1109/ISIT.2007.4557283}}.

\bibitem{BJK11}
Z.~Bar{-}Yossef, Y.~Birk, T.~S. Jayram, T.~Kol, Index coding with side
  information, {IEEE} Transactions on Information Theory 57~(3) (2011)
  1479--1494.
\newblock \href {https://doi.org/10.1109/TIT.2010.2103753}
  {\path{doi:10.1109/TIT.2010.2103753}}.

\bibitem{BKL10}
A.~Blasiak, R.~D. Kleinberg, E.~Lubetzky, Index coding via linear programming,
  CoRR abs/1004.1379 (2010).
\newblock \href {https://doi.org/10.48550/arXiv.1004.1379}
  {\path{doi:10.48550/arXiv.1004.1379}}.

\bibitem{Lee2015}
N.~Lee, A.~G. Dimakis, R.~W. Heath, Index coding with coded side-information,
  IEEE Communications Letters 19~(3) (2015) 319--322.
\newblock \href {https://doi.org/10.1109/LCOMM.2015.2388477}
  {\path{doi:10.1109/LCOMM.2015.2388477}}.

\bibitem{Effros2015}
M.~Effros, S.~El~Rouayheb, M.~Langberg, An equivalence between network coding
  and index coding, IEEE Transactions on Information Theory 61~(5) (2015)
  2478--2487.
\newblock \href {https://doi.org/10.1109/TIT.2015.2414926}
  {\path{doi:10.1109/TIT.2015.2414926}}.

\bibitem{Ong2021}
L.~Ong, B.~N. Vellambi, J.~Kliewer, P.~L. Yeoh, A code and rate equivalence
  between secure network and index coding, IEEE Journal on Selected Areas in
  Information Theory 2~(1) (2021) 106--120.
\newblock \href {https://doi.org/10.1109/JSAIT.2021.3054847}
  {\path{doi:10.1109/JSAIT.2021.3054847}}.

\bibitem{Wan2020}
K.~Wan, D.~Tuninetti, P.~Piantanida, An index coding approach to caching with
  uncoded cache placement, IEEE Transactions on Information Theory 66 (2020)
  1318--1332.
\newblock \href {https://doi.org/10.1109/TIT.2020.2967753}
  {\path{doi:10.1109/TIT.2020.2967753}}.

\bibitem{Dau2012}
S.~H. Dau, V.~Skachek, Y.~M. Chee, On the security of index coding with side
  information, IEEE Transactions on Information Theory 58~(6) (2012)
  3975--3988.
\newblock \href {https://doi.org/10.1109/TIT.2012.2188777}
  {\path{doi:10.1109/TIT.2012.2188777}}.

\bibitem{Narayanan2018}
V.~Narayanan, V.~M. Prabhakaran, J.~Ravi, V.~K. Mishra, B.~K. Dey,
  N.~Karamchandani, Private index coding, in: 2018 IEEE International Symposium
  on Information Theory (ISIT), 2018, pp. 596--600.
\newblock \href {https://doi.org/10.1109/ISIT.2018.8437353}
  {\path{doi:10.1109/ISIT.2018.8437353}}.

\bibitem{Karmoose2017}
M.~{Karmoose}, L.~{Song}, M.~{Cardone}, C.~{Fragouli}, Private broadcasting: An
  index coding approach, in: 2017 IEEE International Symposium on Information
  Theory (ISIT), 2017, pp. 2543--2547.
\newblock \href {https://doi.org/10.1109/ISIT.2017.8006988}
  {\path{doi:10.1109/ISIT.2017.8006988}}.

\bibitem{Mojahedian2017}
M.~M. Mojahedian, M.~R. Aref, A.~Gohari, Perfectly secure index coding, IEEE
  Transactions on Information Theory 63~(11) (2017) 7382--7395.
\newblock \href {https://doi.org/10.1109/TIT.2017.2743688}
  {\path{doi:10.1109/TIT.2017.2743688}}.

\bibitem{Ong2016}
L.~Ong, B.~N. Vellambi, P.~L. Yeoh, J.~Kliewer, J.~Yuan, Secure index coding:
  Existence and construction, 2016 IEEE International Symposium on Information
  Theory (ISIT) (2016) 2834--2838\href
  {https://doi.org/10.1109/ISIT.2016.7541816}
  {\path{doi:10.1109/ISIT.2016.7541816}}.

\bibitem{Cachin1999}
C.~Cachin, S.~Micali, M.~Stadler, Computationally private information retrieval
  with polylogarithmic communication, in: J.~Stern (Ed.), EUROCRYPT '99,
  Springer Berlin Heidelberg, Berlin, Heidelberg, 1999, pp. 402--414.
\newblock \href {https://doi.org/10.1007/3-540-48910-X\_28}
  {\path{doi:10.1007/3-540-48910-X\_28}}.

\bibitem{Chor1997}
B.~Chor, N.~Gilboa, Computationally private information retrieval (extended
  abstract), in: Proceedings of the Twenty-Ninth Annual ACM Symposium on Theory
  of Computing, STOC '97, Association for Computing Machinery, New York, NY,
  USA, 1997, p. 304–313.
\newblock \href {https://doi.org/10.1145/258533.258609}
  {\path{doi:10.1145/258533.258609}}.

\bibitem{Kazemi2019}
F.~Kazemi, E.~Karimi, A.~Heidarzadeh, A.~Sprintson, Single-server
  single-message online private information retrieval with side information,
  in: 2019 IEEE International Symposium on Information Theory (ISIT), 2019, pp.
  350--354.
\newblock \href {https://doi.org/10.1109/ISIT.2019.8849842}
  {\path{doi:10.1109/ISIT.2019.8849842}}.

\bibitem{Heidarzadeh2019}
A.~Heidarzadeh, F.~Kazemi, A.~Sprintson, Capacity of single-server
  single-message private information retrieval with private coded side
  information, in: 2019 IEEE International Symposium on Information Theory
  (ISIT), 2019, pp. 1662--1666.
\newblock \href {https://doi.org/10.1109/ISIT.2019.8849648}
  {\path{doi:10.1109/ISIT.2019.8849648}}.

\bibitem{Heidarzadeh2021}
A.~Heidarzadeh, F.~Kazemi, A.~Sprintson, The role of coded side information in
  single-server private information retrieval, IEEE Transactions on Information
  Theory 67~(1) (2021) 25--44.
\newblock \href {https://doi.org/10.1109/TIT.2020.3029314}
  {\path{doi:10.1109/TIT.2020.3029314}}.

\bibitem{Kushilevitz1997}
E.~Kushilevitz, R.~Ostrovsky, Replication is not needed: single database,
  computationally-private information retrieval, in: Proceedings 38th Annual
  Symposium on Foundations of Computer Science, 1997, pp. 364--373.
\newblock \href {https://doi.org/10.1109/SFCS.1997.646125}
  {\path{doi:10.1109/SFCS.1997.646125}}.

\bibitem{Li2020}
S.~Li, M.~Gastpar, Single-server multi-message private information retrieval
  with side information: the general cases, in: 2020 IEEE International
  Symposium on Information Theory (ISIT), 2020, pp. 1083--1088.
\newblock \href {https://doi.org/10.1109/ISIT44484.2020.9174126}
  {\path{doi:10.1109/ISIT44484.2020.9174126}}.

\bibitem{pid}
H.~Sun, Private information delivery, in: 2019 IEEE Information Theory Workshop
  (ITW), 2019, pp. 1--5.
\newblock \href {https://doi.org/10.1109/ITW44776.2019.8989217}
  {\path{doi:10.1109/ITW44776.2019.8989217}}.

\bibitem{Tandon2017}
R.~Tandon, The capacity of cache aided private information retrieval, in: 2017
  55th Annual Allerton Conference on Communication, Control, and Computing
  (Allerton), 2017, pp. 1078--1082.
\newblock \href {https://doi.org/10.1109/ALLERTON.2017.8262857}
  {\path{doi:10.1109/ALLERTON.2017.8262857}}.

\bibitem{Kadhe2020}
S.~Kadhe, B.~Garcia, A.~Heidarzadeh, S.~El~Rouayheb, A.~Sprintson, Private
  information retrieval with side information, IEEE Transactions on Information
  Theory 66~(4) (2020) 2032--2043.
\newblock \href {https://doi.org/10.1109/TIT.2019.2948845}
  {\path{doi:10.1109/TIT.2019.2948845}}.

\bibitem{Vaidya2023}
K.~Vaidya, B.~{Sundar Rajan}, Cache-aided multi-user private information
  retrieval using {PDA}s, in: 2023 IEEE Information Theory Workshop (ITW),
  2023, pp. 125--130.
\newblock \href {https://doi.org/10.1109/ITW55543.2023.10161671}
  {\path{doi:10.1109/ITW55543.2023.10161671}}.

\bibitem{FiatNaor1994}
A.~Fiat, M.~Naor, Broadcast encryption, in: D.~R. Stinson (Ed.), {CRYPTO} '93,
  Vol. 773 of LNCS, Springer, 1993, pp. 480--491.
\newblock \href {https://doi.org/10.1007/3-540-48329-2\_40}
  {\path{doi:10.1007/3-540-48329-2\_40}}.

\bibitem{GentryW09}
C.~Gentry, B.~Waters, Adaptive security in broadcast encryption systems (with
  short ciphertexts), in: A.~Joux (Ed.), EUROCRYPT 2009, Vol. 5479 of LNCS,
  Springer, 2009, pp. 171--188.
\newblock \href {https://doi.org/10.1007/978-3-642-01001-9\_10}
  {\path{doi:10.1007/978-3-642-01001-9\_10}}.

\bibitem{Agrawal020}
S.~Agrawal, S.~Yamada, Optimal broadcast encryption from pairings and {LWE},
  in: A.~Canteaut, Y.~Ishai (Eds.), EUROCRYPT 2020, Vol. 12105 of LNCS,
  Springer, 2020, pp. 13--43.
\newblock \href {https://doi.org/10.1007/978-3-030-45721-1\_2}
  {\path{doi:10.1007/978-3-030-45721-1\_2}}.

\bibitem{LibertPQ12}
B.~Libert, K.~G. Paterson, E.~A. Quaglia, Anonymous broadcast encryption:
  Adaptive security and efficient constructions in the standard model, in:
  M.~Fischlin, J.~Buchmann, M.~Manulis (Eds.), Public Key Cryptography ({PKC}),
  Vol. 7293 of LNCS, Springer, 2012, pp. 206--224.
\newblock \href {https://doi.org/10.1007/978-3-642-30057-8\_13}
  {\path{doi:10.1007/978-3-642-30057-8\_13}}.

\bibitem{chakraborti2019rOram}
A.~Chakraborti, A.~J. Aviv, S.~G. Choi, T.~Mayberry, D.~S. Roche, R.~Sion,
  {rORAM}: Efficient range {ORAM} with $o(log^2 n)$ locality, in: Symposium on
  the Network and Distributed Systems Security, NDSS'19, ISOC, 2019.
\newblock \href {https://doi.org/10.14722/ndss.2019.23320}
  {\path{doi:10.14722/ndss.2019.23320}}.

\end{thebibliography}

{ \appendix
\label{app}
\section{Rate/Memory Data}
Here we consider the case where $K=N=100$ to quantitatively illustrate the behaviour of the rate/memory curves for various caching schemes.  We see that the rate for uncoded caching is decreasing linearly as $M$ increases.  In contrast, Maddah-Ali and Niesens's coded caching \cite{MaddahAli2014} shows a much steeper initial drop across small values of $M$, displaying the substantial reductions in rate that coded caching can afford. The scheme of Sengupta  et al. \cite{Sengupta2015}, which provides content confidentiality, initially has a higher rate than the unsecured scheme of Maddah-Ali and Niesen, but this also rapidly drops and can be seen to approach the Maddah-Ali and Niesen rate as $M$ becomes larger.  A similar pattern is seen for the demand private scheme of Gurjarpadhye et al. \cite{Gurjarpadhye2021}.  While the the file private scheme of Ravindrakumar et al.\ \cite{Ravindrakumar2016} also displays a steep initial drop in rate, it displays an observably higher rate for large $M$ than the other coded caching schemes.  As discussed in Section~\ref{sec:fileprivacy}, this is to be expected of schemes offering file privacy, for which the rate cannot drop below 1.  
\begin{table}[h!]
\begin{align*}
\begin{array}{rccccc} 
\hline
&\text{Uncoded}& {\text{Maddah-Ali }}&{\text{Sengupta}}&{\text{Ravindrakumar }}&{\text{Gurjurpadhye  }}\\
M&\text{caching}&\text{and Niesen \cite{MaddahAli2014}}&\text{et al.\ \cite{Sengupta2015}}&\text{et al.\ \cite{Ravindrakumar2016}}&\text{et al.\ \cite{Gurjarpadhye2021}}\\
\hline
1	&99&	49.5	&100	&100&	62.76\\
2	&98&	32.67	&49.25	&50.25	&42.13\\
4	&96&	19.2	&24.06	&25.56	&23.47\\
6	&94&	13.43	&15.69	&17.36	&15.58\\
8	&92&	10.22	&11.51	&13.26	&11.47\\
10	&90&	8.18	&9.01	&10.803&8.98\\
12	&88&	6.77	&7.34	&9.17	&7.32\\
14	&86&	5.73	&6.15	&8.00	&6.13\\
16	&84&	4.94	&5.25	&7.12	&5.24\\
18	&82&	4.31	&4.56	&6.44	&4.55\\
20	&80&	3.81	&4.00	&5.89	&4.00\\
22	&78&	3.39	&3.55	&5.45	&3.54\\
24	&76&	3.04	&3.17	&5.08	&3.16\\
26	&74&	2.74	&2.85	&4.76	&2.84\\
28	&72&	2.48	&2.57	&4.49	&2.57\\
30	&70&	2.26	&2.33	&4.26	&2.33\\
40	&60&	1.46	&1.5	&3.44	&1.509\\
50	&50&	0.98	&1	&2.95	&0.99\\
60	&40&	0.66	&0.67	&2.62 &0.67\\
70	&30&	0.42	&0.43	&2.39	&0.43\\
80	&20&	0.25	&0.25	&2.22	&0.25\\
\hline
\end{array}
\end{align*}
\caption{ Approximate rates obtained by caching schemes with $N=K=100$ and memory varying between $1$ and $80$}
\label{tab:quantitative}
\end{table}
}
\end{document}